\documentclass[manuscript]{aastex}
\usepackage{txfonts}
\usepackage{graphicx}

\slugcomment{ApJ, 751, 68 }

\shorttitle{Triggered star formation surrounding Wolf-Rayet star HD
211853} \shortauthors{Liu et al.}

\begin{document}

\title{TRIGGERED STAR FORMATION SURROUNDING WOLF-RAYET STAR HD 211853}

\author{Tie Liu\altaffilmark{1}, Yuefang Wu\altaffilmark{1}, Huawei Zhang\altaffilmark{1}, Sheng-Li Qin\altaffilmark{2}}

\altaffiltext{1}{Department of Astronomy, Peking University, 100871,
Beijing China; liutiepku@gmail.com}
\altaffiltext{2}{I. Physikalisches Institut,
Universit\"at zu K\"oln, Z\"ulpicher Str. 77, 50937 K\"oln, Germany}

\begin{abstract}
The environment surrounding Wolf-Rayet (W-R) star HD 211853 is studied in molecular, infrared, as well as radio,
and Hi emission. The molecular ring consists of well-separated cores, which have a volume density of 10$^{3}$ cm$^{-3}$
and kinematic temperature $\sim$20 K. Most of the cores are under gravitational collapse due to external pressure
from the surrounding ionized gas. From the spectral energy distribution modeling toward the young stellar objects,
the sequential star formation is revealed on a large scale in space spreading from the W-R star to the molecular
ring. A small-scale sequential star formation is revealed toward core "A", which harbors a very young star cluster.
Triggered star formations are thus suggested. The presence of the photodissociation region, the fragmentation of
the molecular ring, the collapse of the cores, and the large-scale sequential star formation indicate that the ¡°collect
and collapse¡± process functions in this region. The star-forming activities in core "A" seem to be affected by the
¡°radiation-driven implosion¡± process.
\end{abstract}


\clearpage

\section{Introduction}
The feedback of massive stars is expected to strongly influence
their surrounding interstellarmedium (ISM) through stellar
radiation, energetic wind, and the ejection of the elemental composition.
The shock front (SF) that emerges as bubbles around
massive stars expand can compress the ISM and trigger star
formation in very dense layers (Chen et al. 2007).

Two models have been proposed to explain the effect of
massive stars on star formation: collect and collapse (Elmegreen
\& Lada 1977) and radiation-driven implosion (RDI; Lefloch
\& Lazareff 1994; Ogura 2010). The former takes place in a
larger spatial size ($\sim$10 pc) with a longer timescale (a few Myr),
whereas the latter takes place in a smaller spatial size ($\sim$1 pc)
with a shorter timescale (0.5 Myr; Ogura 2010). In the "collect
and collapse" scenario, the ionization front (IF) generated from
the Hii region gathers molecular gases and collects a dense
shell between the IF and the SF. The gas and dust in the
collected layer collapse to form stars when they reach the critical
density (Chen et al. 2007; Ogura 2010). This process can selfpropagate
and lead to sequential star formation (Elmegreen \&
Lada 1977). Star formation induced through the "collect and
collapse" process has been revealed in the borders of several
Hii regions (Deharveng et al. 2003, 2005, 2008; Zavagno et al.
2006, 2007; Pomar`es et al. 2009; Petriella et al. 2010;Brand et al.
2011). The so-called RDI process has been proposed to account
for the evolution of bright-rimmed clouds whose surfaces are
ionized by the UV photons from massive stars and trigger the
star formation therein (Bertoldi 1989; Bertoldi \& McKee 1990;
Hester \& Desch 2005; Miao et al. 2009; Gritschneder et al.
2009; Bisbas et al. 2011; Haworth \& Harries. 2012). The IFs
generate a shock that compresses into the cloud and causes
the dense clumps to collapse (Chen et al. 2007). Candidates
for star formation by RDI have been found in many bright rimmed
clouds associated with IRAS point sources (Sugitani
et al. 1991; Sugitani \& Ogura 1994; Urquhart et al. 2004, 2006,
2007; Morgan et al. 2010).

The Wolf.Rayet (W-R) stars, initially more massive than
$\sim$40M$_{\sun}$, have typical mass-loss rates of 10$^{-5}$ M$_{\sun}$~yr$^{-1}$ and
strong stellar winds faster than 1000 km~s$^{-1}$. They can blow
stellar wind bubbles with radii of $\sim$10 to $\sim$100 pc and expanding
velocities of several km~s$^{-1}$ (Marston 1996;Mart\'{\i}n-Pintado et al.
1999). The expanding bubbles can push away the molecular gas
and create molecular layers that fragment into condensations
well separated in due time. The "collect and collapse" process
is expected to happen in the shells swept up by such expanding
stellar wind bubbles (Whitworth et al. 1994). However, the
samples of triggered star formation surrounding W-R stars are
rare and few previous works have focused on star-forming
activities in such regions. In this paper, we provide a sample
region where triggered star formation is witnessed.

HD 211853 is located at the center of an optically bright
region named shell B in the Hii region Sh2-132 (Sharpless 1959;
Harten et al. 1978). The Hii region linked to shell B is excited by
a massive O-type star BD+55.2722 and the W-R starHD211853
(Vasquez et al. 2010). While it is the main ionizing source of the
ring nebula (Cappa et al. 2008), the W-R star HD 211853 (WR
153ab) is a binary system. The spectral types of the two W-R
components are WN6 and WCE+06 (Nishimaki et al. 2008).
The velocity of its stellar wind is as high as 1500.1800 km s.1,
leading to a high mass-loss rate of 10$^{-4.5}$ M$_{\sun}$~yr$^{-1}$ (Nishimaki
et al. 2008). A ring nebula associated with HD 211853 is easily
identified in radio emission, in polycyclic aromatic hydrocarbon
(PAH) emission, and in molecular emission (Vasquez et al. 2010;
Cappa et al. 2008). Figure 1 presents a composite color image
of this region. Blue corresponds to the DSS-R optical emission,
green to the radio emission detected in the Canadian Galactic
Plane Survey (CGPS; Taylor et al. 2003) at 1420 MHz, and red
to 8.3 $\micron$ emission detected at the Midcourse Space Experiment
(MSX) A band.

It can be seen that the radio emission is embraced by
the 8.3 $\micron$ emission, which forms a ring-like structure. A
photodissociated region (PDR) has been identified, and possible
triggered star formation in this region is suggested (Vasquez
et al. 2010). As shown in Figure 2 of Saurin et al. (2010),
three star clusters, "Teutsch 127," "SBB 1," and "SBB 2,"
have been found in this region; the hierarchical structure and
age distribution of the star clusters show evidence of triggered
star formation (Saurin et al. 2010). However, studies of the
star-forming activities in this region are far from mature. In
this paper, through detailed analysis of molecular emission
and characteristics of young stellar objects (YSOs), we have
revealed the picture of triggered star formation in this region.

\section{Observations and Database}
The observations of $^{12}$CO (1-0), $^{13}$CO (1-0), and C$^{18}$O
(1-0) were carried out with the Purple Mountain Observatory (PMO)
13.7 m radio telescope in 2011 January. The new 9-beam array
receiver system in single-sideband (SSB) mode was used at the front
end. The array pixels are sideband separation SIS mixers yielding
18 independent IF outputs. The properties of the Solid-State Imaging Spectrometer (SIS) mixers are introduced in \cite{shan09}. $^{12}$CO (1-0) was observed
at the upper sideband, while $^{13}$CO (1-0) and C$^{18}$O (1-0) were observed simultaneously at the lower
sideband. FFTS spectrometers were used as back ends, which have a
total bandwidth of 1 GHz and 16384 channels, corresponding to a
velocity resolution of 0.16 km~s$^{-1}$ for $^{12}$CO (1-0) and 0.17
km~s$^{-1}$ for $^{13}$CO (1-0) and C$^{18}$O (1-0). The half-power
beam width was 56$\arcsec$, and the main beam efficiency was
$\sim0.5$ during our observation period. The pointing accuracy of the telescope was better than
4$\arcsec$. The typical system temperature (T$_{sys}$) in SSB mode
was around 110 K and varied about 10\% for each beam. The
On-The-Fly (OTF) observing mode was applied. The antenna
continuously scanned a region of 22$\arcmin\times22\arcmin$ centered
on the Wolf-Rayet star HD 211853 with a scan speed of
20$\arcsec$~s$^{-1}$. The OTF data were then converted to three-dimensional cube
data with a grid spacing of 30$\arcsec$. Because the edges of the
OTF maps are very noisy, only the central 14$\arcmin\times14\arcmin$
region was selected to be analyzed. The typical rms noise level was
0.2 K in T$_{A}^{*}$ for $^{12}$CO (1-0), and 0.1 K for $^{13}$CO
(1-0) and C$^{18}$O (1-0). For the data analysis, the GILDAS
software package including CLASS and GREG was employed
\citep{gui00}.

Neutral hydrogen (HI) 21 cm line data were from CGPS with intensities shown as
brightness temperature T$_{mb}$. The synthesized beam and rms bright
temperature in a 0.82 km~s$^{-1}$ channel were $\sim1\arcmin$ and 3
K, respectively. Radio continuum data and images at 1420 MHz were
extracted from the CGPS \citep{tay03} and the NRAO Very Large Array (VLA) Sky Survey
(NVSS, \cite{con98}). The positions and fluxes of the NVSS sources were obtained
from the NVSS catalog \citep{con98}. The CGPS 1420 MHz data had a synthesized beam
of $\sim50\arcsec$ and a noise level of $\sim3$ mJy~beam$^{-1}$. The
NVSS image had a spatial resolution of 45$\arcsec$ and an rms
brightness fluctuation of $\sim0.45$ mJy~beam$^{-1}$ (Stoke I). MSX
A-band data were extracted from the MSX Galactic Plane Survey
\citep{pro01}, which has a spatial resolution of $\sim18\arcsec.3$.
Spitzer IRAC data and Two Micron All Sky Survey (2MASS) data were obtained from \cite{qiu08},
both of which have good spatial resolutions of $\sim2\arcsec$. We
also obtained SCUBA 450 $\micron$ and 850 $\micron$ continuum data
from the SCUBA Legacy Catalogues \citep{di08}. At 850 and 450
$\micron$, the resolutions of SCUBA data are of $\sim14\arcsec$ and
$\sim9\arcsec$, respectively.
\section{Results}

\subsection{Molecular emission}
The $^{12}$CO (1-0), $^{13}$CO (1-0), and C$^{18}$O (1-0) emissions were
observed simultaneously. However we did not detect any C$^{18}$O
(1-0) emission in this region above 3$\sigma$ (0.3 K), which indicates
low column density in this region. $^{12}$CO (1-0) emission was detected in
almost the whole region even at the position of the Wolf-Rayet star
HD 211853. $^{13}$CO (1-0) emission was only detected in the
ring-like region, but not within it. The detailed analysis of the
molecular emission is described below.

\subsubsection{The systemic velocity of the molecular clumps}
In Figure 2, we present the 1$\arcmin$ averaged spectrum of $^{12}$CO (1-0) at
the position of the Wolf-Rayet star. $^{12}$CO (1-0) emission is significantly detected.
From the gaussian fit, the peak antenna temperature is 0.38$\pm$0.02 K,
about 8$\sigma$ level (1$\sigma$ equal to 0.05 K estimated from the
baseline). It has a line width of 4.1$\pm$0.2 km~s$^{-1}$ and a
peak velocity of -43.4$\pm$0.01 km~s$^{-1}$.

The left panel of Figure 3 presents the 14$\arcmin\times14\arcmin$
averaged spectra of the HI 21 cm line, line $^{12}$CO (1-0) and line $^{13}$CO (1-0). 
The right panel of Figure 3 shows the zoomed in averaged spectra of
the $^{12}$CO (1-0) and $^{13}$CO (1-0) lines. There are two components in the $^{12}$CO (1-0) and $^{13}$CO (1-0)
emission, which was also noticed by \cite{vas10}. The narrow one
peaks at -50 km~s$^{-1}$, while the other broader one ranges from
-48 km~s$^{-1}$ to -39 km~s$^{-1}$.

Are both components related to this region? We argue that the
-50 km~s$^{-1}$ maybe a foreground cold cloud, rather than part of this
region based on the following two aspects.

(1). The integrated emission of $^{12}$CO (1-0) in the first two panels of
Figure 5 is concentrated in an isolated and compact core (denoted as core "I")
north-west of HD 211853. We did not detect this component in the
other positions. Additionally, from the P-V diagrams in Figure 6, one
can see that the -50 km~s$^{-1}$ component is also not connected to the
other component in velocity space. \cite{vas10} argued that the $^{12}$CO (1-0) emission between -53.4
and -46.0 km~s$^{-1}$ is also associated with the ring nebula. However, their conclusion may be misleading because their integrated map
is contaminated by the emission between -48 to -46 km~s$^{-1}$. As the channel map shows, the emission in their clouds E and F is dominated
by the emission between -48 and -39 km~s$^{-1}$, rather than the -50 km~s$^{-1}$ component.

(2). The left panel of Figure 4 presents the spectra of the HI 21 cm line, $^{12}$CO (1-0) and $^{13}$CO (1-0)
lines at the peak of core "I". The right panel of Figure 4 shows the zoomed in spectra of the
$^{12}$CO (1-0) and $^{13}$CO (1-0) lines. One can see that the HI emission at the peak of
core "I" has a very narrow absorption dip. Together with the narrow
line width of the molecular emission, an HI Narrow Self-Absorption
(HINSA) system can be identified, suggesting that core "I" is more likely a
foreground cold cloud \citep{li03}. As shown in Table 2, the
kinematic temperature of core "I" (14.6$\pm$0.3 K) is much lower
than that of core "F" (21.7$\pm$1.0 K), indicating that core "I" is more
quiescent than core "F".

In conclusion, the -50 km~s$^{-1}$ molecular component is more
likely from a foreground cold cloud, rather than a part of the
molecular environment surrounding HD 211853. However because the -50 km~s$^{-1}$ molecular component is very near
the ring nebular in the spatial and velocity space, it may also interact with the expanding molecular ring, which needs further
observations to confirm.

From the Gaussian fit, the averaged $^{13}$CO (1-0) spectrum peaks
at -43.6$\pm$0.1 km~s$^{-1}$, very similar to that of the $^{12}$CO (1-0)
emission taken from the central Wolf-Rayet star (-43.4$\pm$0.01
km~s$^{-1}$). Thus, we adopt -43.5 km~s$^{-1}$ as the overall
systemic velocity of the molecular emissions in this region.

\subsubsection{The properties of molecular cores}
Figure 5 displays the channel maps of $^{12}$CO (1-0) emission in contours overlaid on
the 8.3 $\micron$ emission in the MSX A band. The middle velocities of each panel
are plotted on the upper-right corners. In each panel, the integrated velocity interval is 2 km~s$^{-1}$. The
contours are from 20\% to 90\% of the peak intensity in each panel.
It can be seen that the $^{12}$CO (1-0) emission is very clumpy. The
blue shifted gas (-46 km~s$^{-1}$ panel) distributes in the
western and southern regions, while the red shifted gas (-40 km~s$^{-1}$ panel) is mainly in the northern and eastern regions,
indicating a velocity gradient from north-east to south-west. The
gas at -42 and -44 km~s$^{-1}$ forms a ring-like structure harboring several
well separated cores. The ring-like molecular structure perfectly
coincides with the 8.3 $\micron$ emission in the MSX A band.

Figure 6 presents the P-V cuts along two orientations denoted by the
red dashed lines in Figure 7. From the P-V diagrams, it is clear that the -50 km~s$^{-1}$ component is separated with the -43.5
km~s$^{-1}$ component. An amazing aspect of the P-V cuts is that
the relative velocity of the gas increases with the distance to the WR star from
near to far, following a "Hubble law" like distribution, as
denoted by the dashed lines.

Figure 7 displays the integrated intensity map of $^{13}$CO (1-0)
overlaid on the 8.3 $\micron$ emission in gray scale, which is integrated
from -48 to -39 km~s$^{-1}$. The contours are from 10\%
($\sim8\sigma$) to 90\% of the peak emission. Eight cores are
identified and denoted from "A" to "H". The Wolf-Rayet star HD
211853 is drawn as a "cross". IRAS, MSX and AKARI point sources are
marked with "boxes", "triangles," and "stars", respectively. However, one should keep in mind that
many of these point sources may be not real; rather, they may correspond to some
variations of brightness in the infrared emission along the filamentary structures. More observations are needed to constrain
their properties. For convenience, we still call them "point sources" in the following text. These
cores form a ring-like structure surrounding HD 211853, and they are
associated with the 8.3 $\micron$ emission in the MSX A band. Most of the infrared point
sources are distributed in the molecular ring. Core "A" is associated
with an IRAS source "IRAS 22172+5549", while the emission peaks of
the other cores seem to be separated from the infrared point
sources, which will be discussed later in Section 4.3.

Taking into account the effect of beam smearing, the radii of the
cores are calculated as R=$\frac{\sqrt{4A/\pi-\theta_{b}^{2}}}{2}$,
where A is the projected area of each core and $\theta_{b}$ is the
beam width. Adopting a distance of 3 kpc \citep{cap08}, the radii of the cores are
calculated and listed in the fourth column of Table 1. The average
radius is 1 pc.

Figure 8 presents the spectra of $^{12}$CO (1-0), $^{13}$CO (1-0), and
C$^{18}$O (1-0) at the emission peaks of the cores. The spectra at
core "E" are totally blue shifted, while at core "H" the spectra are red-shifted. The
peak velocities of the other cores coincide with or slightly deviate
from the systemic velocity. The results of the Gaussian fit to these
spectra are listed in Table 1. Core "G" has the largest line width
and antenna temperatures, while core "H" has the smallest ones. The
line width of $^{13}$CO (1-0) ranges from 1.3 to 2.6 km~s$^{-1}$
with a mean value of 2.1 km~s$^{-1}$.

Assuming $^{12}$CO (1-0) emission is optically thick and
$^{13}$CO (1-0) is optically thin, we derived the parameters of the
cores including the excited temperatures T$_{ex}$, column densities
of H$_{2}$, N$_{H_{2}}$, and optical depth of $^{13}$CO (1-0) lines
under the local thermal equilibrium (LTE) assumption following
\cite{gar91}. Typical abundance ratios
[H$_{2}$]/[$^{12}$CO]=$10^{4}$ and [$^{12}$CO]/[$^{13}$CO]=60 were
used in the calculations \citep{de08}. The volume density of H$_{2}$ is calculated as
n$_{H_{2}}$=N$_{H_{2}}$/2R. Then the core mass can be derived as
M=$\frac{4}{3}\pi \cdot R^{3}\cdot n_{H_{2}}\cdot m_{H_{2}}\cdot
\mu_{g}$, where $m_{H_{2}}$ is the mass of a hydrogen molecule and
$\mu_{g}$=1.36 is the mean atomic weight of the gas. The derived
parameters are listed from Columns 3-7 in Table 2. The N$_{H_{2}}$
ranges from $1.8\times10^{21}$ to $1.4\times10^{22}$ cm$^{-2}$. The
optical depth of $^{13}$CO (1-0) is around 0.3. The core masses
range from 21 M$_{\sun}$ ("H") to 1061 M$_{\sun}$ ("G") with a mean
value of 400 M$_{\sun}$.

We also applied RADEX \citep{van07}, a one-dimensional non-LTE
radiative transfer code, which uses the escape probability
formulation assuming an isothermal and homogeneous medium without
large-scale velocity fields, to constrain the physical conditions, such as density and kinetic temperature,
of the cores. In the first
running, we explored a range of H$_{2}$ volume densities and
temperatures of [10$^{2}$,10$^{6}$] cm$^{-3}$ and [5,50] K. The
molecular column densities, which are also input parameters
required, are fixed according to the H$_{2}$ volume densities as
N(X)=n$_{H_{2}}\cdot$2R$\cdot$X(x), where X(x) is the relative
abundance ratio of the molecule. For $^{12}$CO (1-0),
X(x)=10$^{-4}$, and for $^{13}$CO (1-0), X(x) is adopted as
1.67$\times10^{-6}$. The model parameters were selected to
satisfy $\mid T_{mod}-T_{obs}\mid\leq3\sigma$, where $T_{mod}$
and $T_{obs}$ are the modeled and observed brightness temperature of
the molecule transition, respectively. As shown in the left panel of
Figure 9, the kinematic temperatures and H$_{2}$ volume densities can
be well constrained by modeling both the $^{12}$CO (1-0) and $^{13}$CO
(1-0) lines. Two geometries, static spherical and expanding
spherical (similar to large velocity gradient (LVG) approximation),
were chosen in this running. It can be seen that both geometry
assumptions give nearly the same results for $^{13}$CO (1-0) lines,
while for $^{12}$CO (1-0) lines, static spherical models give
slightly higher volume densities than expanding spherical at higher
kinematic temperatures. However, in the overlapping region, the
deviations of the two geometries are small enough to be ignored. We
narrowed the parameter space in the second running assuming the
expanding spherical geometry. The range H$_{2}$ volume densities
and temperatures in the second running are
[2.5$\times$10$^{2}$,10$^{4}$] cm$^{-3}$ and [8,30] K, respectively.
As shown in the right panel of Figure 9, the kinematic temperatures and
volume densities of H$_{2}$ can be well constrained in the
overlapping region of $^{12}$CO (1-0) and $^{13}$CO (1-0) tracks.
The best physical parameters from RADEX are obtained by averaging
the parameters in the overlapping region and are displayed from Columns
7-14 in Table 2.

It is interesting to note that the excitation temperatures of
$^{12}$CO (1-0) obtained from RADEX are perfectly consistent with
those calculated under LTE assumption. The excitation temperatures of
$^{13}$CO (1-0) are much lower than those of $^{12}$CO (1-0). The
optical depths of $^{13}$CO (1-0) obtained from RADEX are larger
than those obtained from LTE conditions. The kinematic temperatures
are always larger than the excitation temperatures of $^{12}$CO
(1-0). The volume densities of H$_{2}$ and core masses calculated in
RADEX roughly agree with those from LTE calculations.

\subsection{Radio continuum emission, HI emission and PAH emission}
From the radio emission detected in CGPS at 1420 GHz, \cite{vas10} obtained an rms
n$_{e}$=20 cm$^{-3}$ and M$_{HII}$=1500~M$_{\sun}$ in this region.
As shown in the upper panels of Figure 10, nine NVSS radio sources
(shown in blue dashed contours) are located in this region, and most of
them are bordered by the 8.3 $\micron$ emission ring shown in red solid contours.
The nine radio sources are not individual HII regions, but are instead high brightness components
of the large HII region. They reflect the fluctuation and non-uniformity of the ionized gas distribution
within the large HII region. The Wolf-Rayet star is associated with NVSS source "1", and the
molecular core "A" is associated with NVSS source "5". Following \cite{pan78}, the
electron densities, emission measures and masses of ionized hydrogen
of these NVSS sources are derived by assuming an electronic
temperature of 8000 K. However, those parameters listed in Table 3
can only be treated as lower limits due to the missing flux of VLA.
NVSS "4", "5" and "2" in the yellow dashed box show very high electronic
densities and form an ionization front interacting with the molecular
ring. NVSS "1" and "3" seem to interact with the northern part of
the molecular ring, mainly with core "G".

We integrated HI emission from -48 km~s$^{-1}$ to -39 km~s$^{-1}$
and displayed the integrated intensity map in gray scale in the
upper panel of Figure 10. The HI emission is very
diffuse with an integrated intensity gradient in the N-S orientation. The
HI emission also has a shell-like appearance in the southern part, which is associated with the 8.3 $\micron$ emission.
An HI cavity is found within the yellow dashed box, where the ionization front exists. Assuming that
the HI 21 cm line is optically thin and spin temperature $T_{s}\gg
h\nu/k$, the average column density of HI can be
calculated as $N_{HI} = 1.82\times10^{18}\times\int
\overline{T_{b}}dv$ (cm$^{-2}$). Based on the averaged spectrum of
HI shown in the left panel of Figure 3, the average column density of
HI in the whole region of shell B is about $9.4\times10^{20}$ cm$^{-2}$. The total
mass of HI is estimated to be 1100 M$_{\sun}$, which is comparable to the total mass of HII ($\sim$1500 M$_{\sun}$) \citep{vas10}.

The ring-like feature of the 8.3 $\micron$ emission in MSX A band was
been revealed by \cite{cap08,vas10}. It clearly shown as red
contours in the upper-panel of Figure 10 in this paper. The 8.3
$\micron$ emission is dominated by PAH emission and indicates the
presence of PDRs \citep{vas10}. IRAC 8 $\micron$ emission shown in
color scale in the lower panel of Fig.10 also reveals a shell-like
structure of PAH emission.

In Figure 11, we analyze the distribution of the normalized intensities
of $^{12}$CO (1-0) emission, HI emission, 1.4 GHz radio emission observed
by CGPS and PAH emission revealed by MSX A band along four
orientations centered on the Wolf-Rayet star HD 211853. At first
glance, one can see that $^{12}$CO (1-0) emission and PAH emission encircle a
hollow with an angular radius of $\sim5\arcmin$ in each panel, where
their emissions are suppressed but 1.4 GHz radio
emission is enhanced, indicating that the central region is filled with UV
radiation that has photodissociated and ionized the molecular and
atomic gas. The central Wolf-Rayet star is surrounded by the ionized gas and shows strong
continuum emission at 8.3 $\micron$. The HI emission
always shows distribution opposite to the 1.4 GHz radio emission. In
other words, the peaks and troughs of HI emission always correspond
to the troughs and peaks of the 1.4 GHz radio emission, respectively.
This anti-correlation is more significant in the PA=0$\arcdeg$ plot.
However, we also noticed that the HI emission is also high in some regions
filled with ionized gas, such as in NVSS 1 and 3. This may be caused by
the non-uniform distribution of HI gas, or by the projection of the
atomic gas located at the outer layer of the expanding bubble. The peaks of $^{12}$CO (1-0) emission generally are located outside of PAH emission.
The fact that PAH emission is located at the interface between the
ionized and molecular material suggests that molecular gas is being
photodissociated by the UV photons, i.e., the presence of PDRs
\citep{vas10}. At the radius of $\sim5\arcmin$ in the
PA=135$\arcdeg$ panel, $^{12}$CO (1-0), HI, 1.4 GHz radio and PAH emissions
nearly all reach their maximum emissions. This position corresponds
to molecular core "A", which is a bright Rimmed cloud and is forming
the youngest star cluster in this region \citep{sau10}.

\subsection{Spectral Energy Distributions (SEDs) of YSOs}
\cite{qiu08} identified tens of Young Stellar Objects (YSOs) in the
south-east of this region. We selected 64 of these sources for
analysis. The others are beyond 5$\arcmin$ from the Wolf-Rayet star
HD 211853 and are mostly classified as class II objects. The SEDs of
these YSOs were then modeled using an online two-dimensional radiative transfer
tool developed by \cite{ro06,ro07}. As input parameters, the distances of
these YSOs and the visual extinction A$_{v}$ were explored in a
parameter space of [2.75,3.25] kpc and [2,2.5] mag, respectively.
The adopted input A$_{v}$ is similar to that of HD 211853 (2.28 mag) and BD+55$\arcdeg$2722 (2.26 mag) \citep{vas10}. The
SED models that satisfy the criterion of
$\chi^{2}-\chi_{best}^{2}<3\times n_{data}$, where $n_{data}$ is the
number of data points, were accepted and analyzed. Figure 12 shows the SEDs
of ten stars.

Following \cite{gra09}, a weighted mean and
standard deviation are derived for all the physical parameters
of each source from the selected models, with the weights
being the inverse of the $\chi^{2}$ of each model. The fitting results
are listed in Table 4. The various columns in Table 4 are
as follows: Column 1, the name of the star; Columns 2 and
3, the coordinates of the star; Column 4, the number of
models averaged; Column 5, the weighted $\chi^{2}$; Column 6, the distance of the star; Column 7, the interstellar extinction;
Column 8, the age of the star; Column 9, the mass of the
star; Column 10, the total luminosity of the star; Column 11,
the envelope mass; Column 12, the envelope accretion rate;
Column 13, the disk mass; and Column 14, the disk accretion
rate. The evolutionary stages of these YSOs are defined as
follows (Robitaille et al. 2006): Stage 0/I objects are
those with $\dot{M}_{env}/M_{*}>10^{-6}yr^{-1}$, Stage II objects
are those with $\dot{M}_{env}/M_{*}<10^{-6}yr^{-1}$ and
$M_{disk}/M_{*}>10^{-6}$, and Stage III objects are those with
$\dot{M}_{env}/M_{*}<10^{-6}yr^{-1}$ and $M_{disk}/M_{*}<10^{-6}$.
The identifications of the evolutionary stages are listed in the last column of Table 4.
"Y14" and "Y20" are badly fitted with $\chi^{2}$ much larger than 1000, indicating that
they may not connect to this region. We exclude them in the following analysis. The SED of "Y27", which was once
classified as a class II YSO by \cite{qiu08}, is more likely from the
atmosphere of an evolved star rather than from YSOs. It can be well
modeled by a stellar atmosphere with temperature of 7500 K and
log(g) of 0.5. The last panel of Figure 12 presents its SED. Among
the other YSOs, 40\% of them are younger than 10$^{6}$ yr and 44 have
masses smaller than 2 M$_{\sun}$. The youngest and most massive YSOs
are found associated with molecular core "A".

One should be careful with these results for individual young stars due to the limit of bands used in fitting.
More data, especially at long wavelengths, are required to better constrain the properties of these young stars.
However, the present results can provide us clues as to the age and the mass distribution of these young stars spreading
in this region. To investigate the reliability of the age and mass distribution of these YSOs in the SED fitting, we fit
the SEDs again by loosing the parameter space of A$_{v}$, which is explored in [0,2.5] mag. Figure 13 compares the ages and masses of the YSOs in the two runnings.
One can find that they coincide with each other very well. We therefore believe that even if the parameters of the individual stars need to be further constrained, the trend of the age and mass spatial distribution of these YSOs is reliable in statistics.

\subsection{The properties of core A~/~IRAS 22172+5549}
Molecular core "A" is associated with IRAS 22172+5549, which is an
externally heated rimmed cloud \citep{qiu08}. The lower panel of Figure 10
displays the 850 $\micron$ emission in red solid contours from SCUBA,
which shows a compact dust core. We collected data from
IRAS, AKARI and SCUBA, and modeled the SED with wavelength longer
than 60 $\micron$, whose emission is mainly from a cool dust
envelope, with a simple isothermal gray-body dust model. In the
optically thin case, the gray-body dust model can be written as
(Zhu et al. 2010, and references therein):
\begin{equation}
S_{\nu}\approx\frac{M_{tot}\kappa_{\nu}B_{\nu}(T_{d})}{gD^{2}}
\end{equation}
where S$_{\nu}$ is the continuum emission flux at the frequency
$\nu$, M$_{tot}$ is the total mass of the gas and dust, $\kappa_{\nu}$
is the dust opacity per unit dust mass, g=100 is the density ratio
of the gas to dust, D is the distance, and B$_{\nu}$(T$_{d}$) is the
Planck function with a dust temperature of T$_{d}$.

For a typical dust coagulation timescale $\tau_{coag}=10^{5}$ yr, in the molecular core "A" $n_{H}\times\tau_{coag}\sim10^{8}$ cm$^{-3}$yr$<10^{9}$ cm$^{-3}$yr, indicating that coagulation does not have a considerable effect on the dust opacity in core "A" \citep{oss94}. Thus, the dust opacity of the initial dust distribution (without dust coagulation) with thin ice mantles can be used in the SED fitting \citep{oss94}. The dust opacity at wavelength $\lambda$ can be derived as $\kappa_{\nu}=0.797\times(\frac{\lambda}{1000
\micron})^{-1.86}$, which is from fitting the data in Table 1 of \cite{oss94}. Figure 14 shows the SED of IRAS 22172+5549 overlaid by the best fit curve.
The dust temperature from the best fit is 26.0$\pm$0.6 K, which is larger than the
kinematic temperature derived from the CO emission, suggesting the dust
emission traces a more innermost part than CO gas. The total mass of
gas and dust traced by dust emission is 102$\pm$13 M$_{\sun}$, which
is much smaller than that traced by CO gas. The bolometric luminosity integrated
from 20 to 3000 $\micron$ is 2200 L$_{\sun}$. The dust temperature and the bolometric
luminosity obtained here are consistent with the dust temperature (28$\pm$5 K) and IR luminosity (2700 L$_{\sun}$) calculated using only the measured
flux densities at 60 and 100 $\micron$ \citep{vas10}.

An embedded infrared cluster "SBB1" is found in this molecular core
\citep{sau10}. From SED modeling, there are found to be associated with this molecular core nine YSOs younger than
$5\times10^{5}$ yr, three of which
are even younger than $5\times10^{4}$ yr, indicating that "SBB1" is a
very young star cluster. As shown in Figure 12, all nine YSOs have large infrared
excesses. The three most massive ones are "Y54", "Y55" and "Y56", which have stellar mass $\geq6$ M$_{\sun}$.
The other six sources have stellar mass  $\leq3$ M$_{\sun}$. The infall
rate of these nine YSOs is found larger than $\sim1\times10^{-5}~M_{\sun}\cdot yr^{-1}$. "Y57" has the largest infall rate of
$\sim1\times10^{-3}~M_{\sun}\cdot yr^{-1}$. All rates indicate that young stars are forming in core "A".

\section{Discussion}
\subsection{The geometry of the ring nebular}
The geometry of the ring nebular is important to understanding the interaction between the central Wolf-Rayet star and its environment. In this section, we argue that the molecular ring surrounding HD 211853 is more likely flattened rather than spherical. First, there is less molecular emission within the interior of the molecular ring, which excludes the spherical shell hypothesis. Second, from the first moment map of $^{12}$CO (1-0) in the left panel of Figure 15, one can find that the velocity distribution of the molecular gas is not spherical. The gas is redshifted and blueshifted in the north-east and south-west, respectively. But the gas along the NW-SE direction is much less shifted, indicating that the molecular gas is distributed in a ring with a thickness much smaller than the size of the bubble. The plane in which the molecular ring lies has an inclination angle with respect to the sky plane, leading to the velocity gradient of the molecular gas along NE-SW direction. Finally, as shown in Figure 2, the spectrum of $^{12}$CO (1-0) at the position of the WR star only has one velocity component, which excludes the two expanding layer hypothesis. The geometry and the formation of molecular rings surrounding the expanding bubbles were discussed in detail by \cite{bea10}. In their interpretation, the morphology of molecular rings naturally arises if the host molecular clouds are oblate, even sheetlike, with thicknesses of a few parsecs. The central massive stars can clear out a cavity in the flattened molecular cloud and create the molecular ring surrounding them. This is only in the case of the molecular ring surrounding HD 211853.

A simple geometric model is sketched in the right panel of Figure 15. The axes x and y lie on the plane of the sky. The WR star HD 211853 is located at the origin. The ring nebular is shown as a red circle. The projection of the ring nebular in the sky plane is shown as a blue ellipse. The lengths of the semimajor and semiminor axis of the ellipse are denoted as "a" and "b". The angle $\theta$ is the inclination angle between the plane of the ring nebular and the sky plane. The angle $\phi$ is the position angle (P.A.) measured from the major axis. From this simple geometric model, the inclination angle $\theta$ can be inferred from:
\begin{equation}
\cos \theta=\frac{b}{a}
\end{equation}
The observed velocity, $V_{obs}$ of a point on the ellipse is given by:
\begin{equation}
V_{obs}=V_{exp}\cdot\delta\cdot\sin\theta\sqrt{\frac{1}{1+\cos^{2}\theta\cdot\cot^{2}\phi}}+V_{sys}
\end{equation}
where $V_{sys}$ is the systemic velocity, $V_{exp}$ is the expanding velocity, $\delta$=1 when $0\arcdeg\leq\phi\leq180\arcdeg$ and $\delta$=-1 when $180\arcdeg\leq\phi\leq360\arcdeg$. The two dashed ellipses in the left panel of Figure 15 approximately mark the  boundary of the interior and exterior of the molecular ring. The inner and outer ellipses are of $4\arcmin.5\times3\arcmin$ and $7\arcmin\times5\arcmin$ in diameter. Thus the inclination angle between the plane of the ring nebular and the sky plane turns out to be $\sim44\arcdeg-48\arcdeg$. From the left panel of Figure 15, the maximum value of $\mid V_{obs}-V_{sys}\mid$ is found to be 3.5 km~s$^{-1}$ along the direction of the minor axis, indicating an expanding velocity of $\sim5$ km~s$^{-1}$.

As shown in Figure 1, the ring nebular surrounding HD 211853 is part (shell B) of the large HII region Sh2-132. Although shell B is mainly ionized and created by HD 211853, its interaction with the other part (shell A) cannot be ignored. The formation of shell B should be affected by the expansion of shell A, especially in the west, where the molecular ring is not closed. In the above model, we did not take into account the interaction with shell A, which needs more detailed modeling. However, the simple model can roughly depict the ring structure surrounding HD 211853 and explain the spatial and velocity distribution of the molecular gas in the ring nebular.

\subsection{Gravitational stability of the molecular cores}
The gravitational stability of the molecular cores can be investigated by
comparing their core masses with virial masses and Jeans masses.
Assuming that the cloud core is a gravitationally bound isothermal sphere
with uniform density and is supported solely by random motions, the
virial mass M$_{vir}$ can be calculated following \cite{ung00} as following:
\begin{equation}
\frac{M_{vir}}{M_{\sun}}=2.10\times10^{2}(\frac{R}{pc})(\frac{\Delta V}{km~s^{-1}})
\end{equation}
where R is the radius of the core and $\Delta V$ is the line width of $^{13}$CO (1-0).
The virial masses are listed in the 15th column of Table 2.

Many factors, including thermal pressure, turbulence, and magnetic
field, support the gas against gravity collapse in molecular cores.
Taking into account the thermal and turbulent support, the Jeans mass
can be expressed as \citep{hen08}:
\begin{equation}
M_{J}=a_{J}\frac{C_{s,eff}^{3}}{\sqrt{G^{3}\rho}}
\end{equation}
where $a_{J}$ is a dimensionless parameter of order unity which
takes into account the geometrical factor, $\rho$ is the mass
density and $C_{s,eff}$ is an effective sound speed including
turbulent support,
\begin{equation}
C_{s,eff}=[(C_{s})^{2}+(\sigma_{NT})^{2}]^{1/2}
\end{equation}
where $C_{s}$ is the thermal sound speed and $\sigma_{NT}$ is the non-thermal one dimensional velocity dispersion. The thermal sound speed
is related with the kinematic temperature as following: $C_{s}=(kT_{k}/\mu_{g}
m_{H_{2}})^{1/2}$, where $\mu_{g}$=1.36 is the mean atomic weight of the gas. The non-thermal one dimensional velocity dispersion $\sigma_{NT}$ can be calculated as follows:
\begin{equation}
\sigma_{NT} =\sqrt{\sigma_{^{13}CO}^{2}-\frac{kT_{k}}{m_{^{13}CO}}}
\end{equation}
and
\begin{equation}
\sigma_{^{13}CO}=\frac{\Delta V}{\sqrt{8ln(2)}}
\end{equation}
with $m_{^{13}CO}$ being the mass of $^{13}CO$.
By introducing an effective kinematic temperature $T_{eff}=\frac{C_{s,eff}^{2} \mu_{g} m_{H_{2}}}{k}$,
equation (5) can be rewritten in a form similar to Equation (18) in
\cite{hen08},
\begin{equation}
M_{J}\approx1.0a_{J}(\frac{T_{eff}}{10~K})^{3/2}(\frac{\mu}{2.33})^{-1/2}(\frac{n}{10^4~cm^{-3}})^{-1/2}M_{\sun}
\end{equation}
where n is the volume density of H$_{2}$ calculated with RADEX, and $\mu$ is the mean molecular weight of the gas. The Jeans masses
calculated are listed in the 16th column of Table 2.

The core masses of cores "C", "E", and "H" are found to be much smaller than their viral masses and Jeans masses, while the core masses of the other cores
are comparable with their virial masses and Jeans masses. Keeping this in mind, we did not consider the external pressure from their ionized boundary layers in calculating the virial masses and Jeans masses above. It seems that these cores are gravitationally stable against collapse without external pressure. However, the rising external pressure surrounding them can significantly change their stabilities and induce collapse in them \citep{tho04}. The external pressure from the ionized layer is \citep{mor04}:
\begin{equation}
\frac{P_{ext}}{k}=2\rho_{i}C_{i}^{2}=4n_{e}T_{e}
\end{equation}
The effective electron temperature $T_{e}$ is assumed to be 8000 K. The electron densities $n_{e}$ in the ionized boundary layers surrounding cores "A", "B", and "G" are taken as the values of NVSS sources 5, 4, and 1, respectively. For the other cores, we assume $n_{e}$ to be 20 cm$^{-3}$ \citep{vas10}. The calculated external pressure from the ionized layers can be found in the 18th column in Table 2. Molecular pressure inside the clouds can be expressed as:
\begin{equation}
\frac{P_{mol}}{k}=nT_{eff}
\end{equation}
The inferred $P_{mol}$ can be found in the last column of Table 2. We can see for all the cores that external pressure from the surrounding ionized gas is much greater than the molecular pressure, indicating that the photoionization-induced shocks can penetrate the interiors of the molecular cores and compress materials inside \citep{mor04}. To explore the stabilities of these cores subject to non-negligible external pressure, we calculated the "pressurized virial mass" as \citep{tho04}:
\begin{equation}
M_{pv}\simeq5.8\times10^{-2}\frac{(\Delta V)^{4}}{G^{3/2}P_{ext}^{1/2}}
\end{equation}
where G is the gravitational constant and $\Delta V$ is the line width of $^{13}$CO (1-0) lines. The calculated "pressurised virial masses" are listed in the 17th column of Table 2. The molecular cores with core masses exceeding their "pressurized virial masses" are unstable against gravitational collapse \citep{tho04}. Core "C", "E" and "H" have core masses smaller than their "pressurized virial masses", indicating they are stable even under external pressure. For the other cores, their core masses are much larger than their "pressurized virial masses", suggesting that they are likely to be unstable against collapse. It can be seen the high external pressure has a great destabilizing effect upon the gravitational stabilities of the cores.

\subsection{The effect of the Wolf-Rayet star on the molecular distribution}
It is expected that the current WR stars and their previous main sequence
phases should greatly influence their associated clouds.
\cite{vas10} has demonstrated that the mechanical energy released by
the Wolf-Rayet star HD 211853 can shape the ring-like structure of
the molecular and PAH distributions.

From Equation (3), one can see that the gases distributed along a direction of a given P.A. should have the same observed velocity. However,
this is not true in the first moment map in the left panel of Figure 15. Instead, one can notice that the expanding velocity increases with the distance from the origin, especially for the gases along the NE-SW direction. This situation can also be found in the P-V diagrams in Figure 6, where the velocity distribution shows a "Hubble law". Such a velocity gradient was also noticed in the
molecular clouds surrounding O stars \citep{dent09}. It seems that the
expansion of the ionized gas can constantly accelerate the molecular
gas. However, the acceleration is not significant toward the dense
cores as shown in Figure 8. The dense molecular cores seem to have
slight velocity deviation from the systemic velocity, which can be
explained by the fact that the expansion of the ionized gas is decelerated by the
resistance of the dense cores due to their large column densities.
HD 211853 has a significant effect on its surrounding
molecular gas by sweeping, photodissociating, reshaping and possibly
collecting.

We also noticed that the infrared point sources in the molecular ring are
always separated from the molecular cores. This
effect is more significant toward cores "C" and "F", where the
infrared sources are located in front of the molecular core and face the
central Wolf-Rayet system, indicating that the molecular cores have
different expanding velocities with the infrared point sources.
This separation indicates that stellar wind has different effects on the dust and gas.
However, such separation maybe caused by the separation of the
formed stars from their parent clouds during evolution or by
inducing new cores by those formed stars. Detailed analysis and
observations are needed to settle this problem.

\subsection{Sequential star formation in the vicinity of core "A"}
\subsubsection{Large-Scale sequential star formation}
YSOs with various ages are drawn with different markers in Figure 16
overlaid on the column density of H$_{2}$ in gray scale and excitation
temperature of $^{12}$CO (1-0) in contours. The interesting thing is
that the age gradient of the YSOs appears in a large-scale. The
oldest YSOs with ages larger than $2\times10^{6}$ yr are scatted in a large
area from the Wolf-Rayet system to the molecular ring. The younger
ones with ages greater than 5$\times10^{5}$ yr but smaller than
10$^{6}$ yr are located in the midway, while the youngest ones are found
associated with molecular core "A". Such an age gradient can also be identified
in panel (i) of Figure 2 in \cite{qiu08}. In their paper, the evolutionary stages
of the YSOs in this region are identified in color-color diagrams.

We averaged the ages of the YSOs in each 0$\arcmin$.5
radius bin and present the radial age distribution in Figure 17. The projection effect was not taken into account.
Since the YSOs are mainly distributed in the vicinity of core "A", the projection effect should not greatly affect
the trend of the age distribution. We find that the YSOs
within 2$\arcmin$ from the Wolf-Rayet system are as old as
2$\times10^{6}$ yr. Those between 2$\arcmin$ and 4$\arcmin$ are aged $\sim1.2\times10^{6}$ yr. The youngest stars are beyond
4$\arcmin$ and have an average age of $\sim2\times10^{5}$ yr. It is
clear that the nearer YSOs are older than the farther ones,
indicating a stellar age gradient. The stellar age distribution
suggests that at least three generations of stars have been forming in
this region. The oldest generation is closer to the Wolf-Rayet
system, while the youngest one is located in the molecular ring. In addition, the three star clusters in this region have different ages \citep{sau10}.
The "Teutsch 127", which includes the O-type star BD+55$\arcdeg$2722 and near the Wolf-Rayet star HD 211853, is the oldest one ($\sim$5 Myr). The "SBB 2"
is as old as $\sim$2 Myr. The age of the youngest cluster "SBB 1" is about 1 Myr. The age gradient of these three star cluster also indicates
sequential star formation on a large scale. Such
"relay star formation" hints at trigged star formation due to
the expansion of bubbles created by stellar wind of massive stars
\citep{chen07}.

\subsubsection{Small-Scale sequential star formation}
As shown in Figure 16, a small-scale age gradient is found to be associated
with molecular core "A," a bright rimmed cloud. The older and less
massive YSOs embrace the molecular core, while the youngest and more
massive stars reside deeper in the core. This picture is in
accordance with the so-called "radiation-driven implosion" (RDI)
process as depicted in Figure 1 of \cite{og06}, which is proposed for triggered star formation in a bright
rimmed cloud \citep{ber89,ber90}. In the model, the surface layer of
the cloud is ionized by the UV photons from massive stars and then
the cloud is compressed and collapsed due to the shock generated
from the ionization front. As discussed in Section 4.2, core "A" is likely to collapse
due to external pressure from the ionized boundary layer. Recent simulation suggests that there is a range of ionizing fluxes in the RDI model,
that trigger star formation in the bright rimmed cloud \citep{bis11}. This range is about $10^{9}$cm$^{-2}$s$^{-1}\lesssim\Phi_{LyC}\lesssim3\times10^{11}$cm$^{-2}$s$^{-1}$. If the ionizing flux is larger
than $10^{9}$cm$^{-2}$s$^{-1}$, the shock front preceding the D-type ionization front can propagate into the cloud, compress it, and trigger star formation therein \citep{bis11}. However large ionizing flux can rapidly disperse the cloud without triggering star formation \citep{bis11}. The ionizing fluxes surrounding core "A" can be estimated as \citep{mor04}:
\begin{equation}
\Phi_{LyC}=1.24\times10^{10}S_{\nu}T_{e}^{0.35}\nu^{0.1}\theta^{-2}
\end{equation}
where $S_{\nu}$ is the integrated radio flux in mJy, $T_{e}$ is the effective
electron temperature of the ionized gas in K, $\nu$ is the frequency in GHz and $\theta$ is the angular diameter
over which the emission is integrated in arcseconds. Assuming $T_{e}=8000$ K, the ionizing flux in NVSS 5 is inferred as $1.1\times10^{10}$cm$^{-2}$s$^{-1}$, which is taken as the value in the ionizing boundary layer surrounding core "A". This value is moderate, indicating that triggered star formation can take place in core "A". In the simulation of \cite{bis11}, however, the RDI model favors the formation of low mass stars in the clouds, which contradicts the situation in core "A", where YSOs with masses larger than 6 M$_{\sun}$ are found. We noticed that the initial cloud mass in their simulation is much smaller than the mass of the core "A", which may be responsible for the contradiction. Additionally, other processes such as "collect and collapse" may also play important roles in star formation in core "A", which will be discussed in the next section.

\subsection{Collect and Collapse scenario}
The ring-like structure of dust and gas distribution and the
existence of PDR indicate that the star formation in this region favors
the "collect and collapse" scenario \citep{de05,vas10}. The molecular
ring is composed of well separated, dense and massive cores, most of which are
collapsing. Such regular spacing rule of the cores could not be
pre-existing clumps but are formed by external triggering, which is a strong evidence for the "collect
and collapse" process \citep{de05,og10}. This situation contradicts the RDI model, in which the dense clumps into which the radiatively induced
shocks drive are pre-existing. Additionally, the large scale sequential star formation discussed in Section 4.4.1 also favors the "collect and collapse" process rather than the "RDI" model \citep{el77,og10}.

However, as discussed in Section 4.4.2, the bright rimmed nebular core "A" also seems to favor the "RDI" model. The "collect and
collapse" process and the "Radiation Driven Implosion" process cannot take place at the same time and place. To resolve this contradiction, we propose that these two processes may function at different times in this region. First, neutral materials are collected and accumulated into a dense shell between the ionization front (IF) and shock front (SF) generated by the expanding HII region. Second, the compressed shocked layer becomes gravitationally unstable and fragments into regularly spaced massive clumps on a long timescale. These two steps mirror the "collect and collapse" process. Third, the clumps are illuminated and eroded by the ionization front, leading to the formation of bright rims and/or cometary globules that collapse to form stars due to the external pressure as discussed in Section 4.2. This third step is like the "RDI" process. Besides, the ionization front illuminating the bright rimmed nebular core "A" is more likely generated from young cluster "SBB 2" rather than from the central Wolf-Rayet star \citep{sau10}. Relatively speaking, core "A" is pre-existing compared with the ionization front generated from the older cluster. In addition, a possible bow-shock seems to have been generated in the cluster "SBB 2" and seems to have an impact on "SBB 1" \citep{sau10}. We therefore suggest that the star formation in core "A" may also be affected by the older cluster "SBB 2" through a process like "RDI", leading to the formation of the youngest embedded star cluster "SBB 1".

\subsection{Comparison with the other star forming regions}
It is necessary to compare the characteristics of the molecular cores in
this region with those of the other star forming regions. The line width of
$^{13}$CO (1-0) ranges from 1.3 to 2.6 km~s$^{-1}$ with a mean value
of 2.1 km~s$^{-1}$, which is very similar to that of the
intermediate-mass star forming regions ($\sim$2 km~s$^{-1}$)
\citep{liu11,sun06}. It is larger than that of the low-mass star forming
regions ($\sim$1.3 km~s$^{-1}$) \citep{my83}, but much smaller than
that of the high-mass star forming regions associated with IRAS sources
($>$ 3 km~s$^{-1}$) \citep{wang09,wu03,wu01} or methanal masers
($\sim$4.5 km~s$^{-1}$) \citep{liu10}. Judging from the line width
of $^{13}$CO (1-0) in this region, the molecular cores where the
triggered star formations possibly take place favor formation of
intermediate-mass YSOs, which is confirmed by SED modeling of the
YSOs in molecular core A. As discussed in Section 4.4.2, core A is
suggested to suffer an RDI process. A
tendency toward HAeBe stars forming deeper in the cloud is expected
in RDI model \citep{og06,chen07}. \cite{liu11} found that the
regions harboring HAeBe stars younger than 10$^{6}$ yr have an
average column density of $\sim5\times10^{21}$ cm$^{-2}$,
slightly smaller than that in most cores of this region. They also have an
average excitation temperature of $\sim$16 K in those HAeBe regions
younger than 10$^{6}$ yr, which is slightly lower than that of the
molecular cores of this region ($\sim$20 K), indicating the
influence of external heating.

\section{Summary}
We studied the environment surrounding Wolf-Rayet star HD 211853 in
molecular, infrared, radio and HI
emission. The SEDs of the YSOs that are located in the south-east
part of this region are modeled with an online YSO SED fitting
tool. The triggered star formation is discussed. The main findings of
this paper are as follows:

(1). The molecular emission has two components. One peaks at $-50$
km~s$^{-1}$, which is likely from a foreground cold cloud. The other
one has a systemic velocity of $-43.5$ km~s$^{-1}$, whose $^{13}$CO
(1-0) emission forms a ring-like structure with several well
separated molecular cores. The line width of $^{13}$CO (1-0) ranges
from 1.3 to 2.6 km~s$^{-1}$ with a mean value of 2.1 km~s$^{-1}$,
which is very similar to that of intermediate-mass star forming
regions.

(2). The properties of the molecular cores are analyzed under the LTE
assumption as well as with non-LTE models in RADEX. Both LTE and
non-LTE analysis give the same excitation temperatures for $^{12}$CO
(1-0) ($\sim$20 K). The excitation temperatures of $^{13}$CO (1-0)
calculated with RADEX are much smaller than those of $^{12}$CO (1-0).
The core masses calculated under LTE and non-LTE are similar,
ranging from $\sim10^{2}$ to $\sim10^{3}$ M$_{\sun}$. The volume densities
of H$_{2}$ in this cores are $\sim10^{3}$ cm$^{-3}$.

(3). The molecular emission is associated with 8.3$\micron$ emission
in the MSX A band. We also witnessed structures like cavities and shells
in HI emission. Radio emission at 1.4 GHz is embraced by the
molecular and dust ring. Ionization fronts easily can be identified
between the radio and molecular emission especially in the
south-east part.

(4). We find that the molecular gas surrounding HD 211853 is not spherically expanding. The molecular gas distributes in a ring-like
structure with a thickness much smaller than the
size of the bubble. The inclination angle between the plane of the ring nebular and the sky plane turns out to be $\sim44\arcdeg-48\arcdeg$. The expanding velocity of the ring nebular is about $\sim5$ km~s$^{-1}$. The expanding velocity seems to increase with the distance
from the WR star.

(5). The molecular cores are well separated along the ring. It seems that these
cores are gravitationally stable against collapse without external pressure. However the high external pressure from the ionized gas has a great destabilizing effect
upon the gravitational stabilities of the cores. Core "C", "E" and "H" have core masses smaller than their "pressurized
virial masses", indicating that they are stable even under external pressure. For the other cores, their
core masses are much larger than their "pressurized virial masses", suggesting that they are likely to be
unstable against collapse.

(5). From the SED modeling of the YSOs located in the south-east of
HD 211853, a stellar age gradient is found across a large scale in
space from the Wolf-Rayet star to the molecular ring. The presence of
PDR, the fragmentation of the molecular ring, the collapse of the cores, and the large scale sequential star
formation indicate that the "collect and collapse" process can work in this region.

(6). Molecular core "A" is forming a star cluster as young as
$\sim10^{5}$ yr. Core "A" is embraced by the ionization front and
older less massive YSOs. More massive YSOs are forming deeper in
core A. The ionizing flux from the ionized boundary layer surrounding core
"A" is moderate enough to generate a leading shock, which propagates into the cloud, compresses it, and triggers
star formation therein. This is very similar to the picture depicted by the "Radiation-Driven Implosion" models.

In conclusion, star formation in the molecular ring is induced
by the Wolf-Rayet star HD 211853 through the interaction between the expanding
ionizing gas and the ISM. The evolution of the molecular ring and the triggered star formation may be affected by both the "collect and collapse" and the "Radiation-Driven Implosion" processes. Generally speaking, the materials in the host molecular cloud surrounding the WR star are collected into a molecular ring. Then the ring fragments into well separated cores that favor the "collect and collapse" process. The cores are unstable due to the external pressure and collapse inside to form stars. The ionization front proceeded by a shock front generated from the older star cluster (like "SBB 2") can propagate into the cloud (core "A") and compress it and trigger star formations therein as depicted by the "radiation-driven implosion" process.

\section*{Acknowledgment}
\begin{acknowledgements}
We are grateful to the staff at the Qinghai Station and Key Laboratory for Radio Astronomy of PMO, CAS, for their assistance
during the observations. This work was partly supported by the NSFC under grants No. 11073003,
10733030, and 10873019, and by the National Key Basic Research
Program (NKBRP) No. 2007CB815403 and No. 2012CB821800.
\end{acknowledgements}

\clearpage

\begin{figure}
\includegraphics[angle=0,scale=.50]{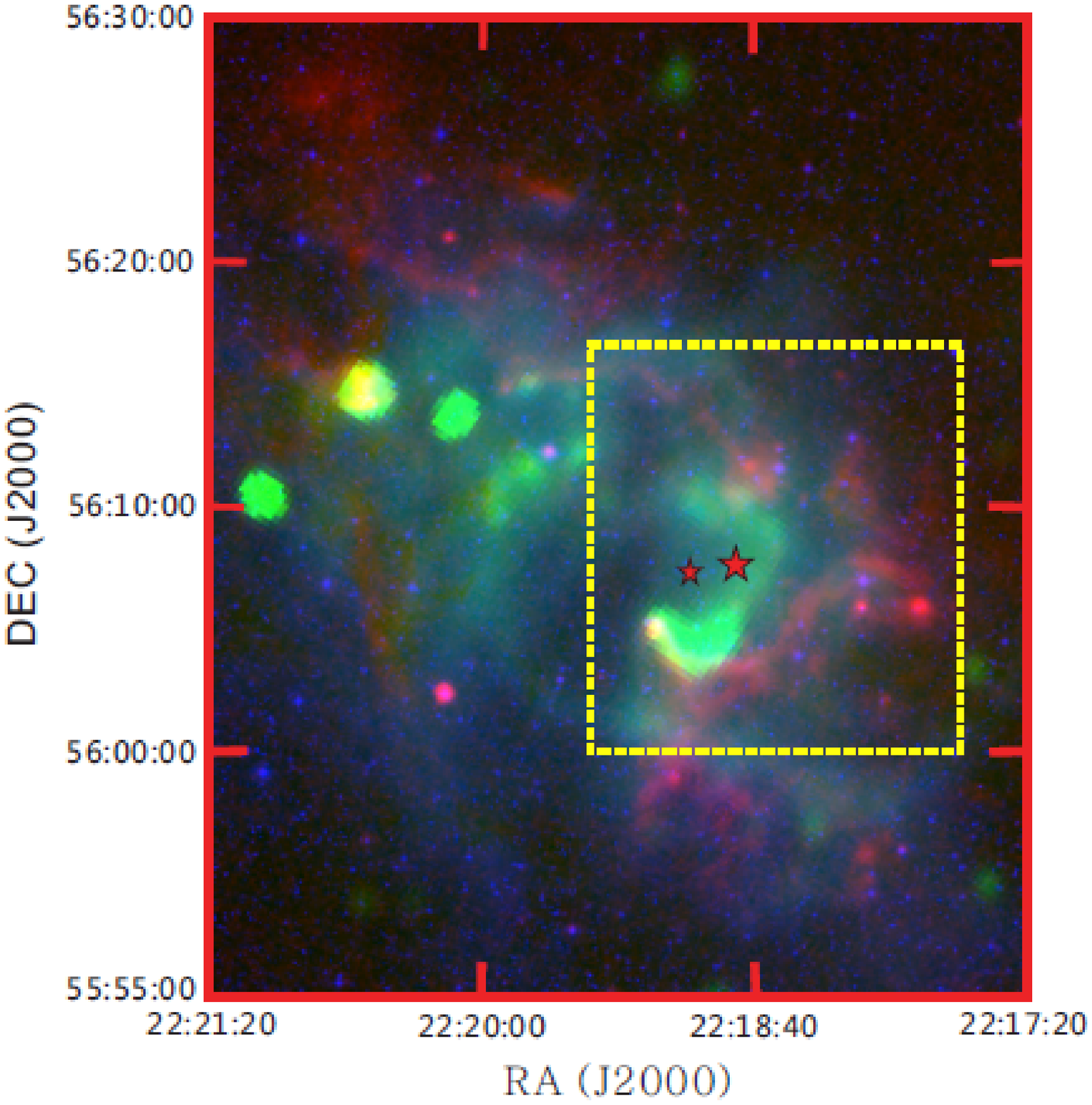}
\caption{Composite color image in this region: DSS2-R optical
emission is blue, 1420 GHz radio emission detected in CGPS is green,
and 8.3 $\micron$ emission detected at MSX A band is red. The larger "star" shows the location of
the WR star HD 211853 and the small "star" represents the position of the massive O-type star BD+55$\arcdeg$2722.
The yellow dashed "box" shows the region of shell B, on which we focus in this paper.}
\end{figure}

\begin{figure}
\includegraphics[angle=90,scale=.50]{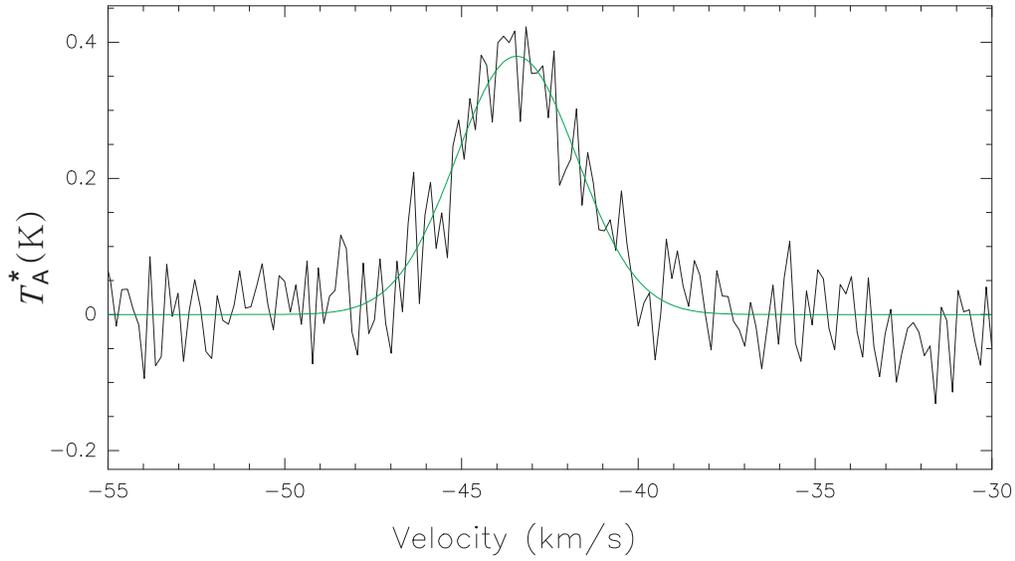}
\caption{Spectrum of $^{12}$CO (1-0) at the position of the Wolf-Rayet star
HD 211853 averaged over 1$\arcmin$. The green solid line is the
Gaussian fit curve.}
\end{figure}

\begin{figure}
\begin{minipage}[c]{0.5\textwidth}
  \centering
  \includegraphics[width=80mm,height=65mm,angle=0]{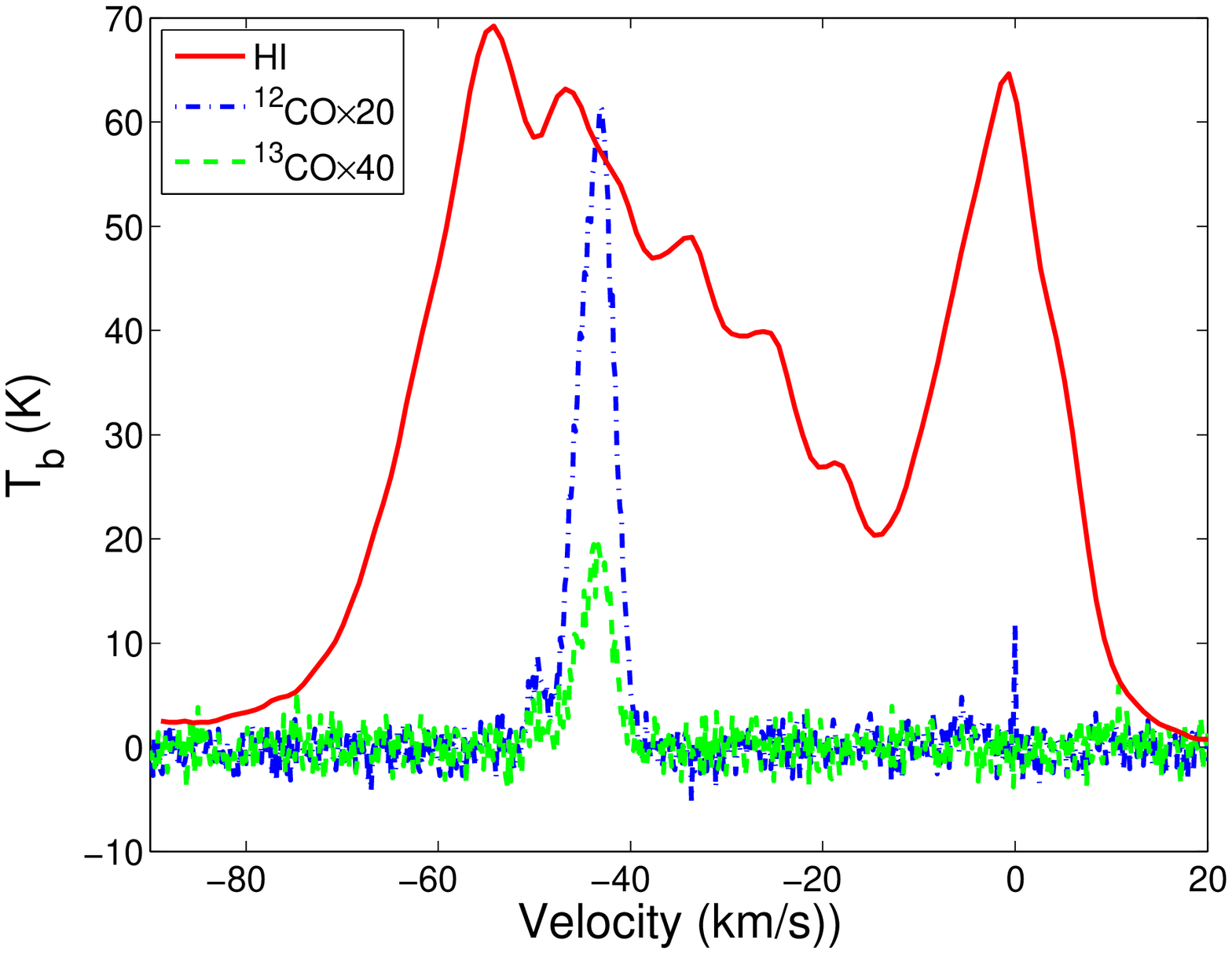}
\end{minipage}
\begin{minipage}[c]{0.5\textwidth}
  \centering
  \includegraphics[width=80mm,height=65mm,angle=0]{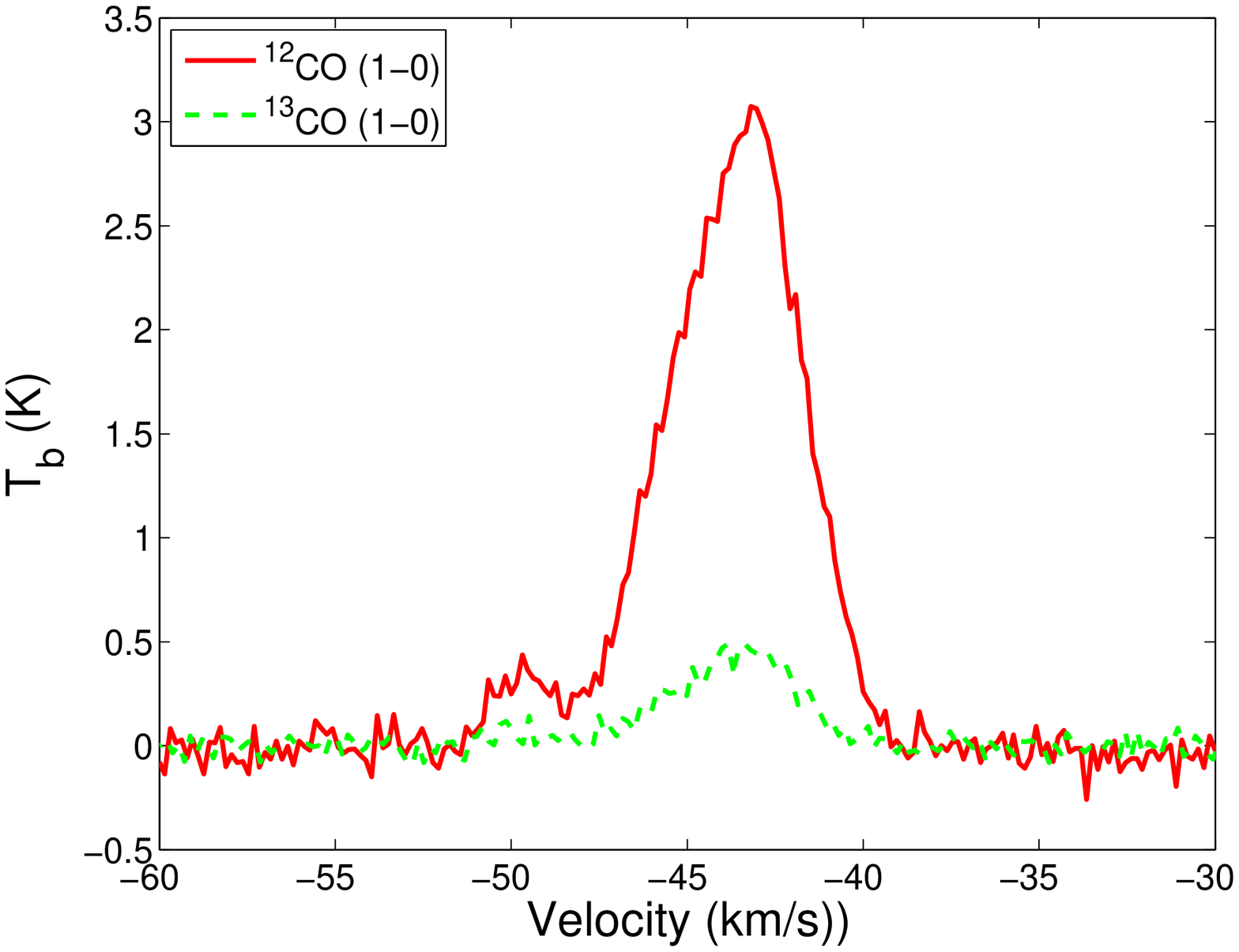}
\end{minipage}
\caption{Left: Spectra of $^{12}$CO (1-0), $^{13}$CO (1-0) and HI averaged over an area
of 14$\arcmin\times14\arcmin$ centered at the Wolf-Rayet star HD
211853. Right: Zoomed in averaged spectra of $^{12}$CO (1-0), and $^{13}$CO (1-0).}
\end{figure}

\begin{figure}
\begin{minipage}[c]{0.5\textwidth}
  \centering
  \includegraphics[width=80mm,height=65mm,angle=0]{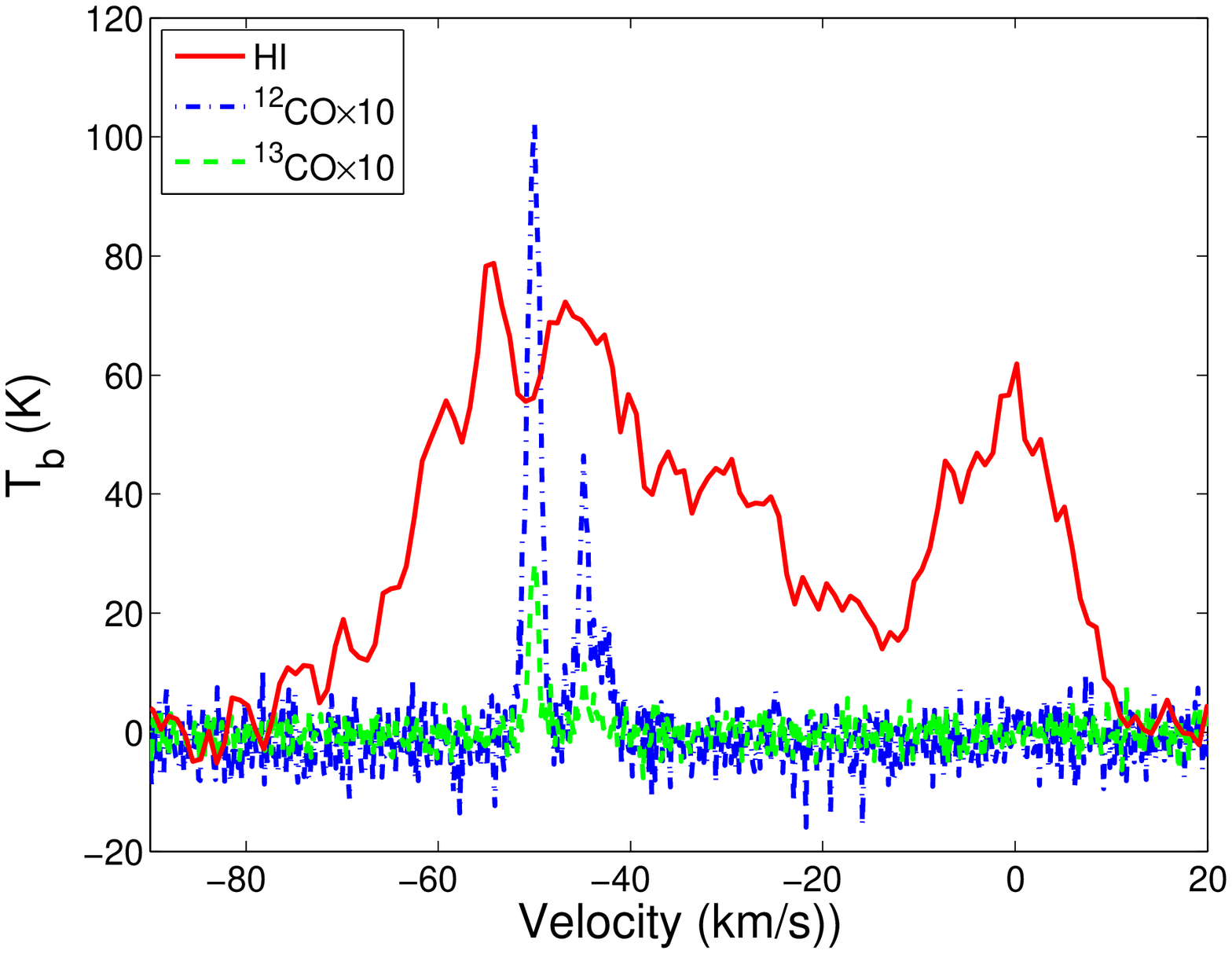}
\end{minipage}
\begin{minipage}[c]{0.5\textwidth}
  \centering
  \includegraphics[width=80mm,height=65mm,angle=0]{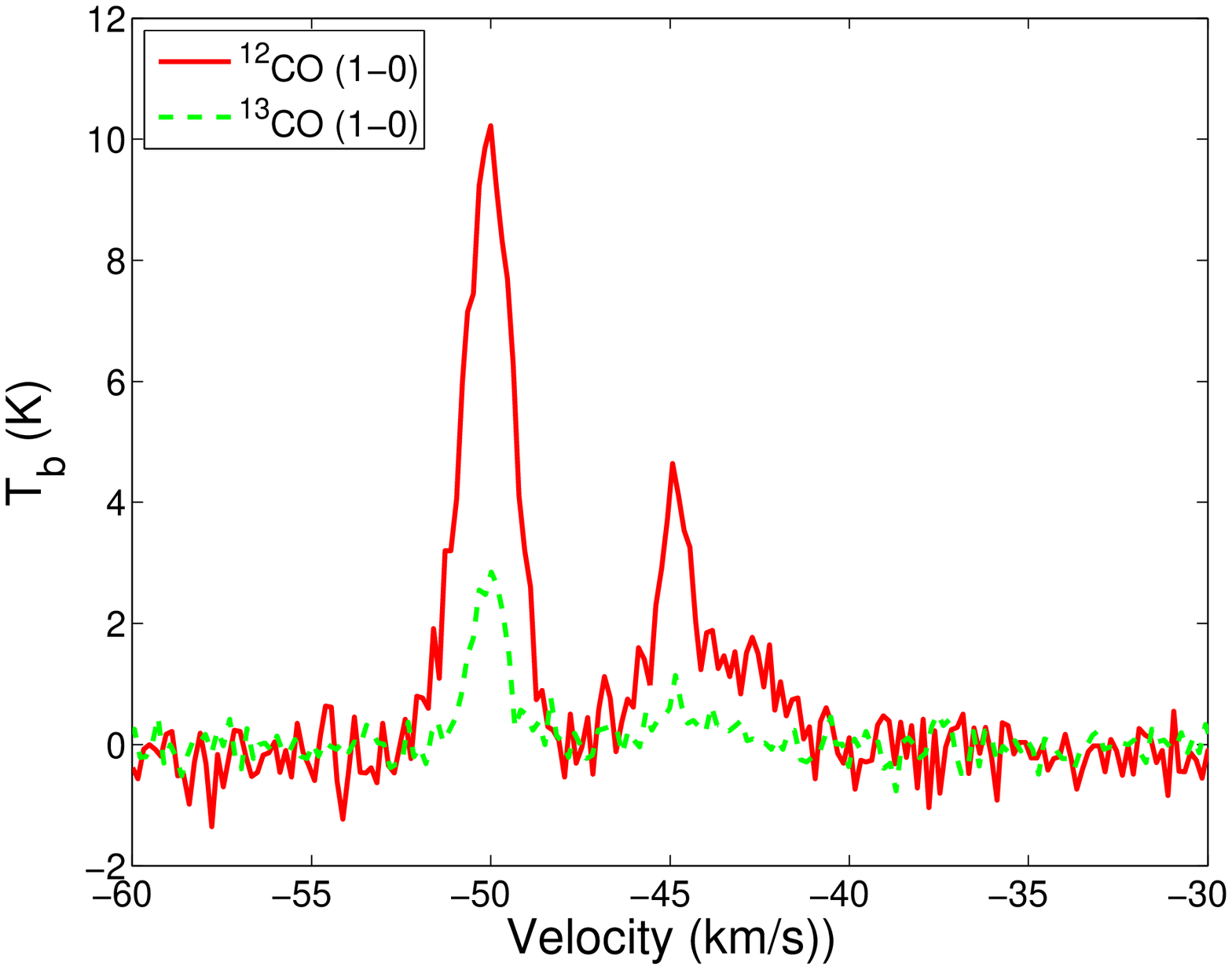}
\end{minipage}
\caption{Left: Spectra of $^{12}$CO (1-0), $^{13}$CO (1-0) and HI taken from the
emission peak of the component at -50 km/s (core "I"). Right: Zoomed in spectra of $^{12}$CO (1-0), and $^{13}$CO (1-0) at core "I".}
\end{figure}

\begin{figure}
\includegraphics[angle=-90,scale=.50]{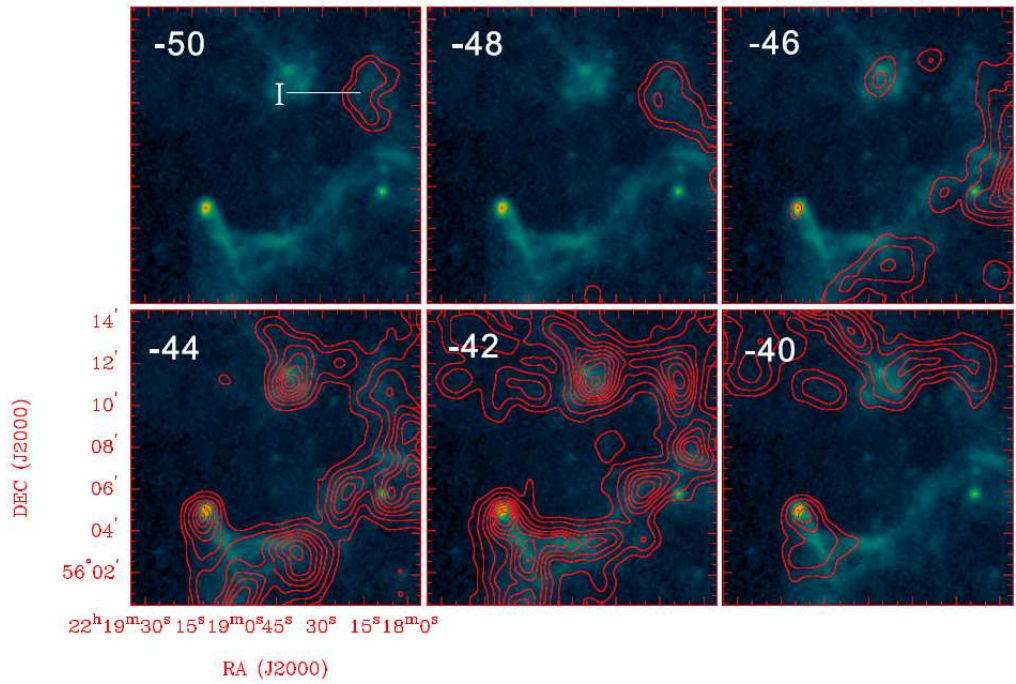}
\caption{Channel maps of $^{12}$CO (1-0). The middle velocities are plotted on the
upper-right corners of each panel. The integrated velocity interval
in each panel is 2 km~s$^{-1}$. The contours are from 20\% to 90\% of the peak
intensity (4.6 K$\cdot$km~s$^{-1}$). The background image shows the 8.3 $\micron$ emission detected at the MSX A band. The position of core "I"
is marked in the -50 km~s$^{-1}$ panel.}
\end{figure}

\begin{figure}
\includegraphics[angle=0,scale=.50]{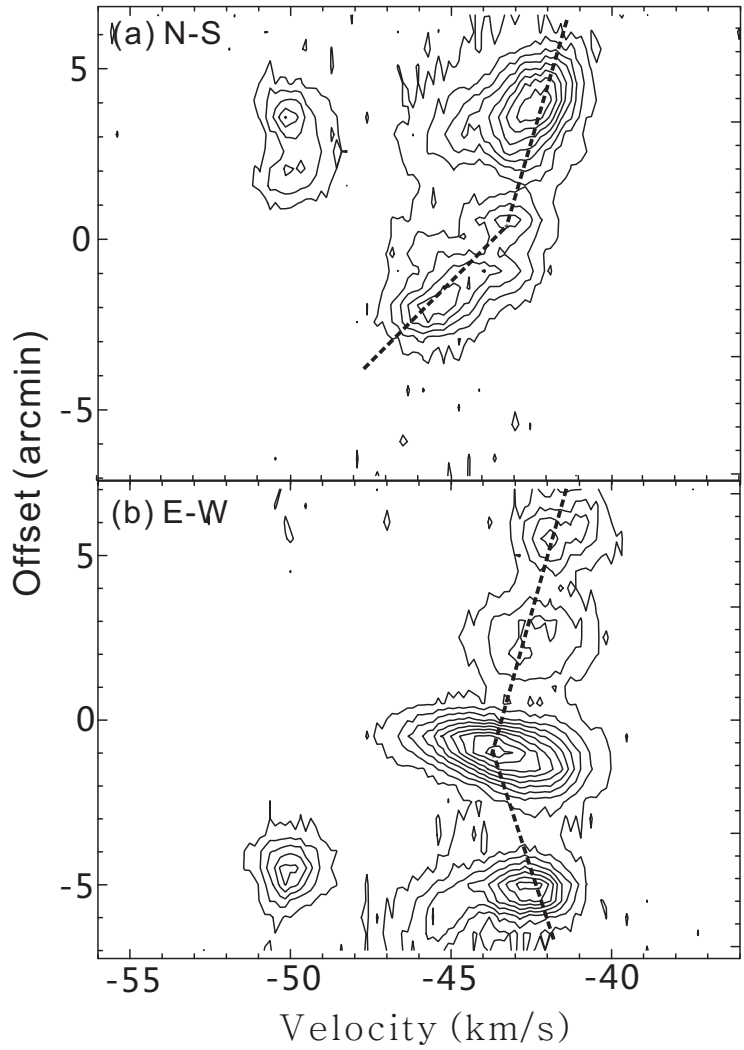}
\caption{P-V diagrams of $^{12}$CO (1-0) cut along the orientations
denoted in Figure 7. The contours are from 0.6 K (3$\sigma$) in steps of 1 K.}
\end{figure}

\begin{figure}
\includegraphics[angle=0,scale=.50]{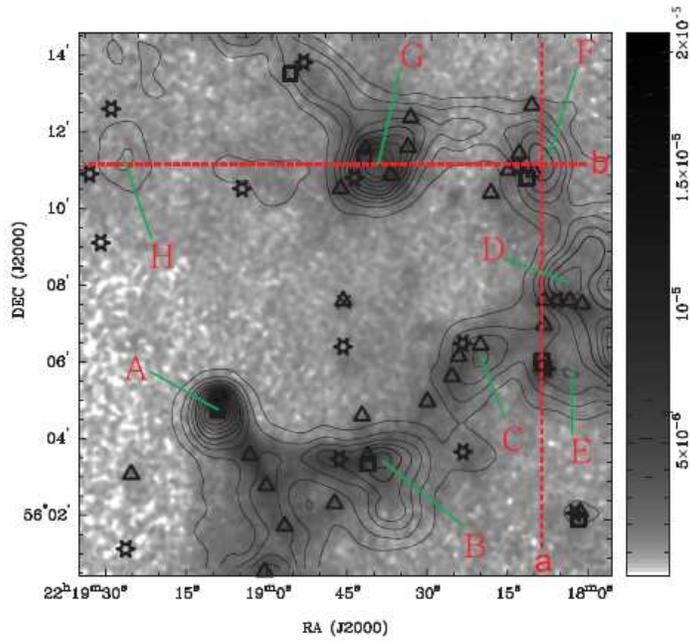}
\caption{$^{13}$CO (1-0) integrated intensity map overlayed on MSX A band
emission. The contours are from 10\% to 90\% of the peak (7.8
K$\cdot$ km~s$^{-1}$). "Triangles" represent MSX point sources,
"boxes" IRAS point sources, and "stars" AKARI point sources. The
Wolf-Rayet star is drawn as "cross". The molecular cores are denoted
from "A" to "H". The horizon and vertical dashed lines represent
the directions of the P-V cuts in Figure 6.}
\end{figure}

\begin{figure}
\includegraphics[angle=0,scale=.50]{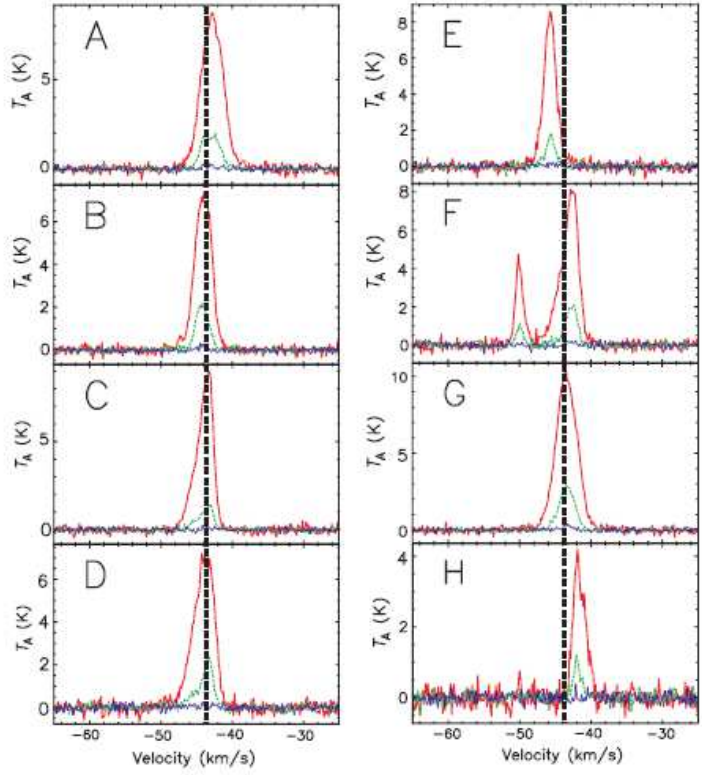}
\caption{Spectra of $^{12}$CO (1-0) are red, those of $^{13}$CO (1-0) are green
and those of C$^{18}$O (1-0) are blue. The core names are labeled on the
upper-left corners of each panel. The dashed black lines indicate
the systemic velocity of -43.5 km~s$^{-1}$. }
\end{figure}

\begin{figure}
\begin{minipage}[c]{0.5\textwidth}
  \centering
  \includegraphics[width=80mm,height=65mm,angle=0]{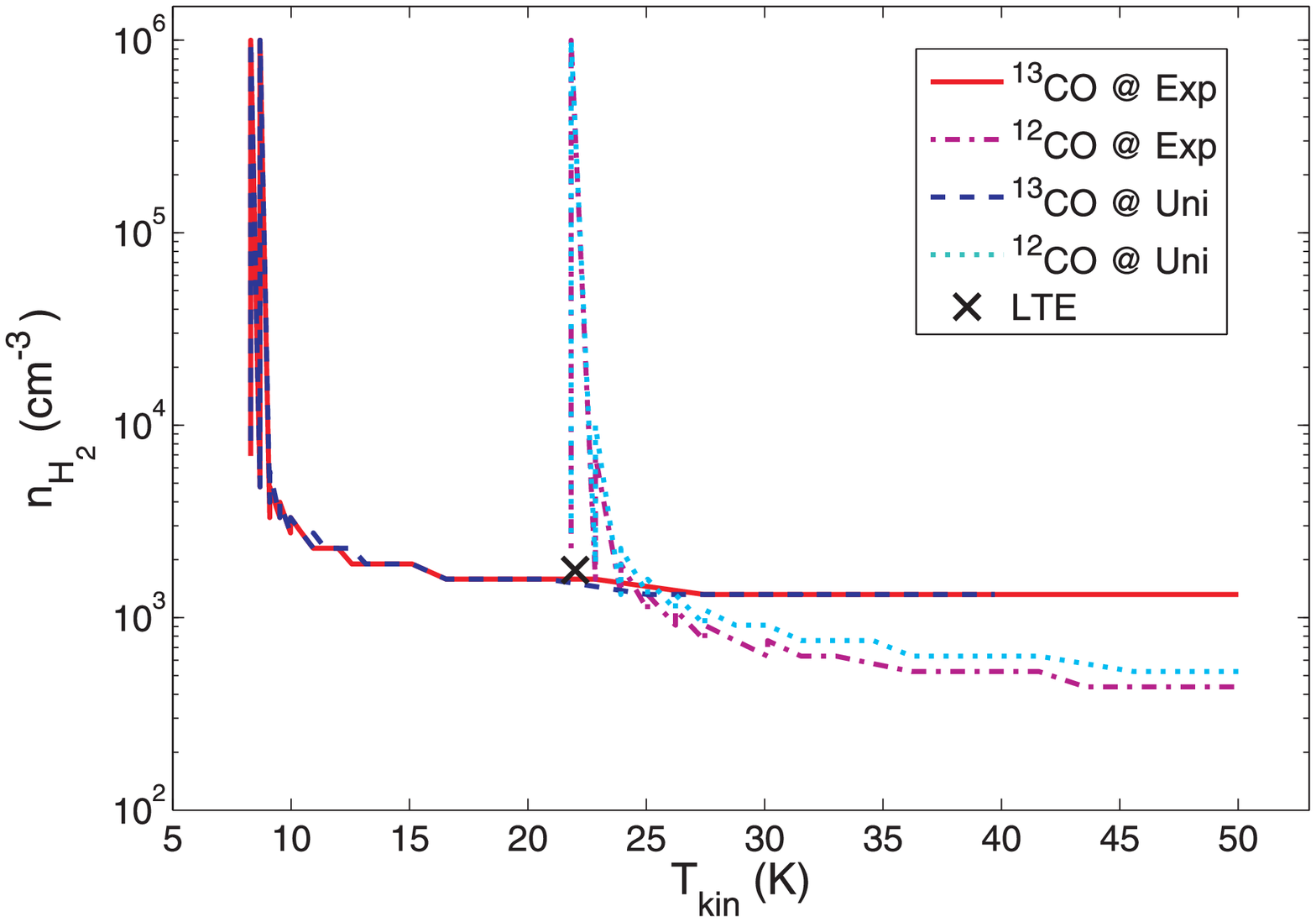}
\end{minipage}
\begin{minipage}[c]{0.5\textwidth}
  \centering
  \includegraphics[width=80mm,height=65mm,angle=0]{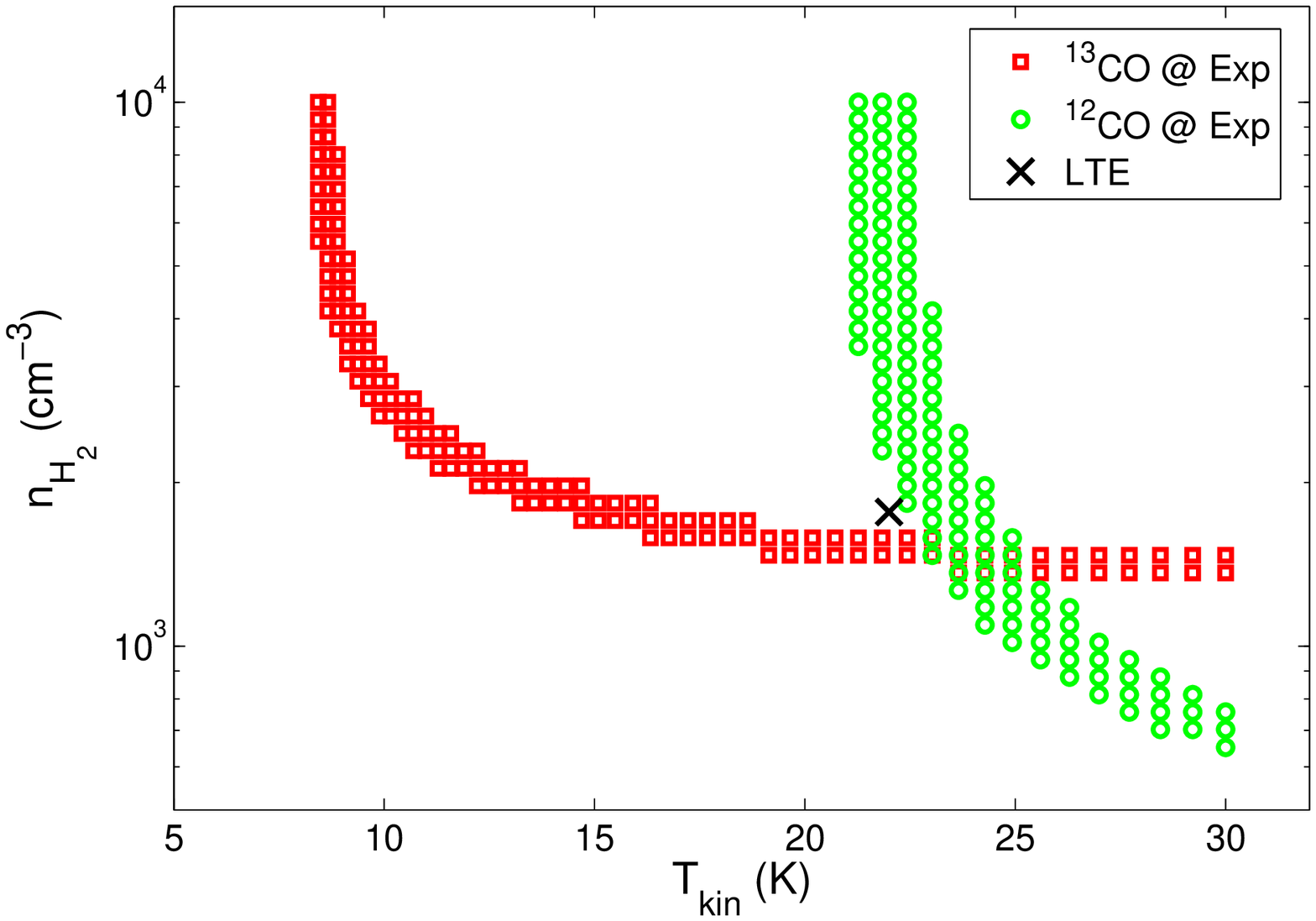}
\end{minipage}
\caption{Non-LTE model fitting of $^{12}$CO (1-0) and $^{13}$CO
(1-0) at the emission peak of core "A". Left: parameters run through
T$_{kin}$ of [5,50] K and n$_{H_{2}}$ of [10$^{2}$,10$^{6}$]
cm$^{-3}$ in log space. Right: parameters run through T$_{kin}$ of
[8,30] K and n$_{H_{2}}$ of [2.5$\times$10$^{2}$,10$^{4}$] cm$^{-3}$
in log space.}
\end{figure}

\begin{figure}
\includegraphics[angle=0,scale=.50]{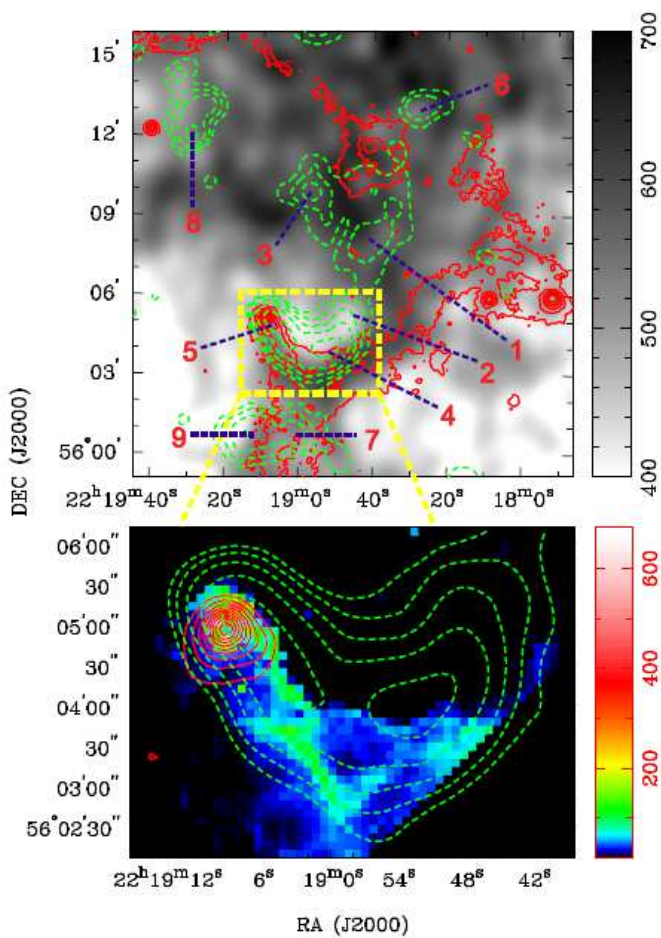}
\caption{Upper: NVSS 1.4 GHz emission is shown in green dashed contours (0.45
mJy (1 $\sigma$) $\times$ 3,6,9,...), MSX A band emission is shown in red
solid contours (from 10\% to 90\%), and HI emission integrated from -48
km~s$^{-1}$ to -39 km~s$^{-1}$ is in gray scale. Lower: NVSS 1.4
GHz emission is shown in green dashed contours, SCUBA 850 $\micron$ emission is
in red solid contours (from 10\% to 90\%) and IRAC 8 $\micron$ emission is
in color scale.}
\end{figure}

\begin{figure}
\begin{minipage}[c]{0.5\textwidth}
  \centering
  \includegraphics[width=80mm,height=65mm,angle=0]{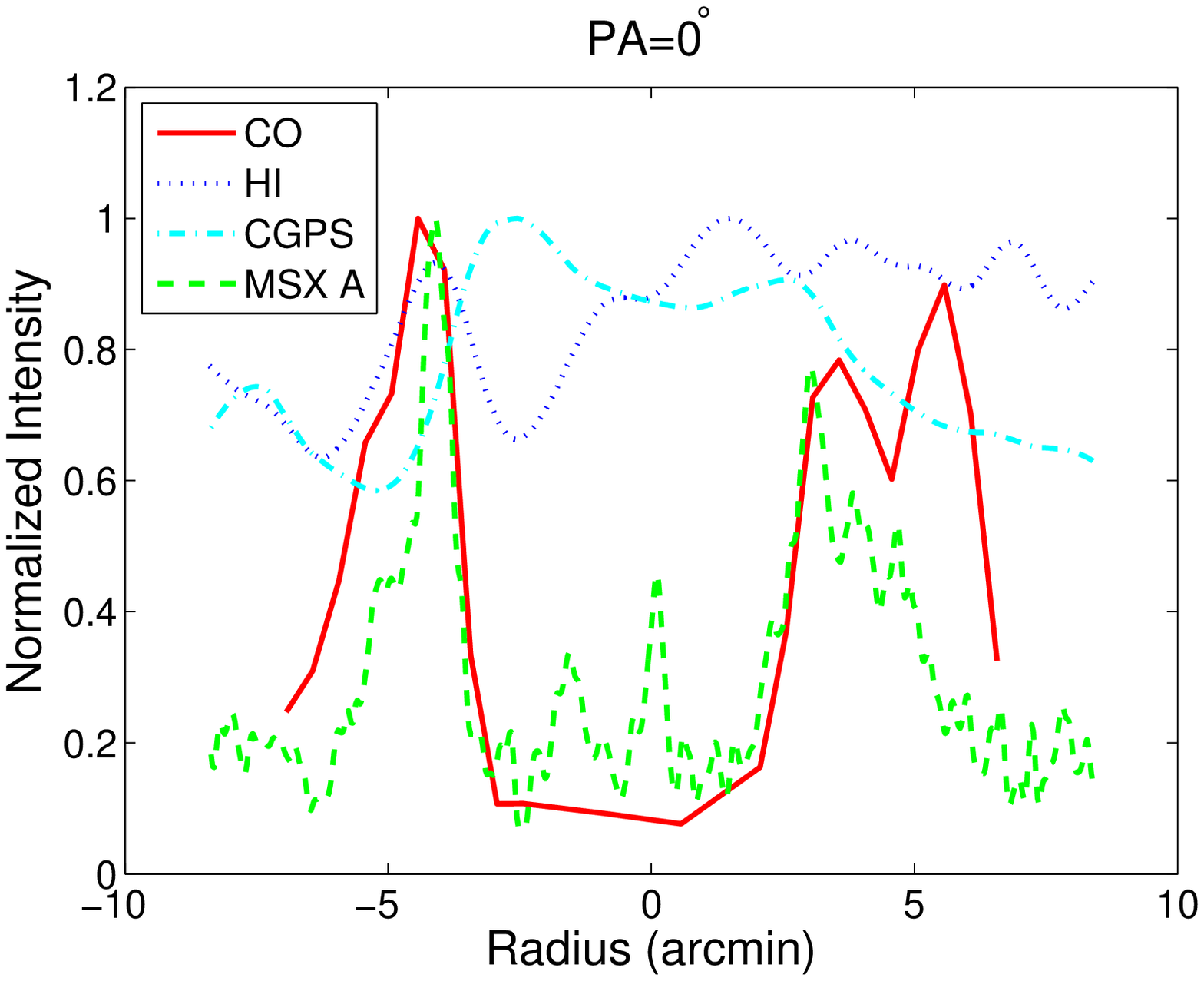}
\end{minipage}
\begin{minipage}[c]{0.5\textwidth}
  \centering
  \includegraphics[width=80mm,height=65mm,angle=0]{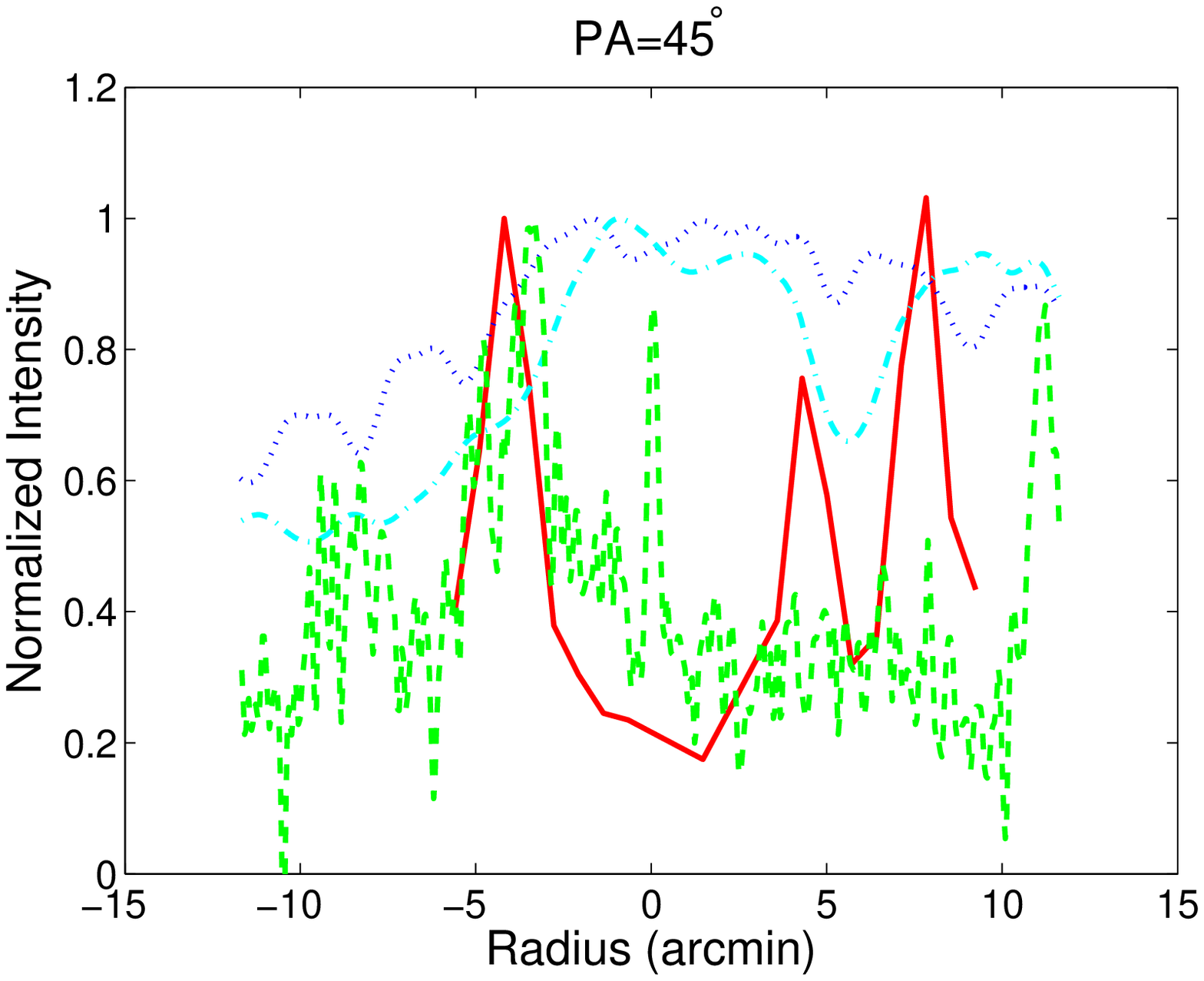}
\end{minipage}
\begin{minipage}[c]{0.5\textwidth}
  \centering
  \includegraphics[width=80mm,height=65mm,angle=0]{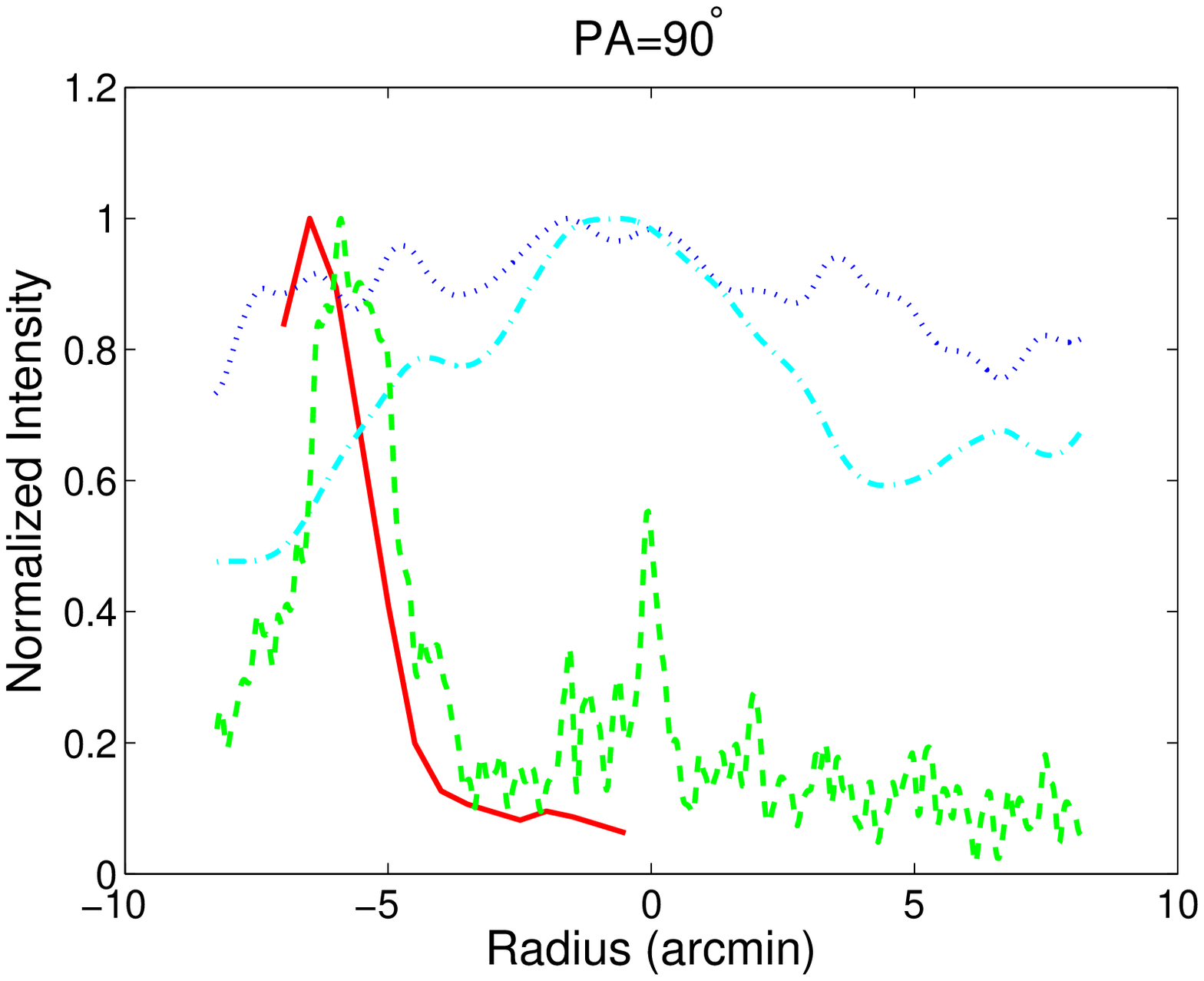}
\end{minipage}
\begin{minipage}[c]{0.5\textwidth}
  \centering
  \includegraphics[width=80mm,height=65mm,angle=0]{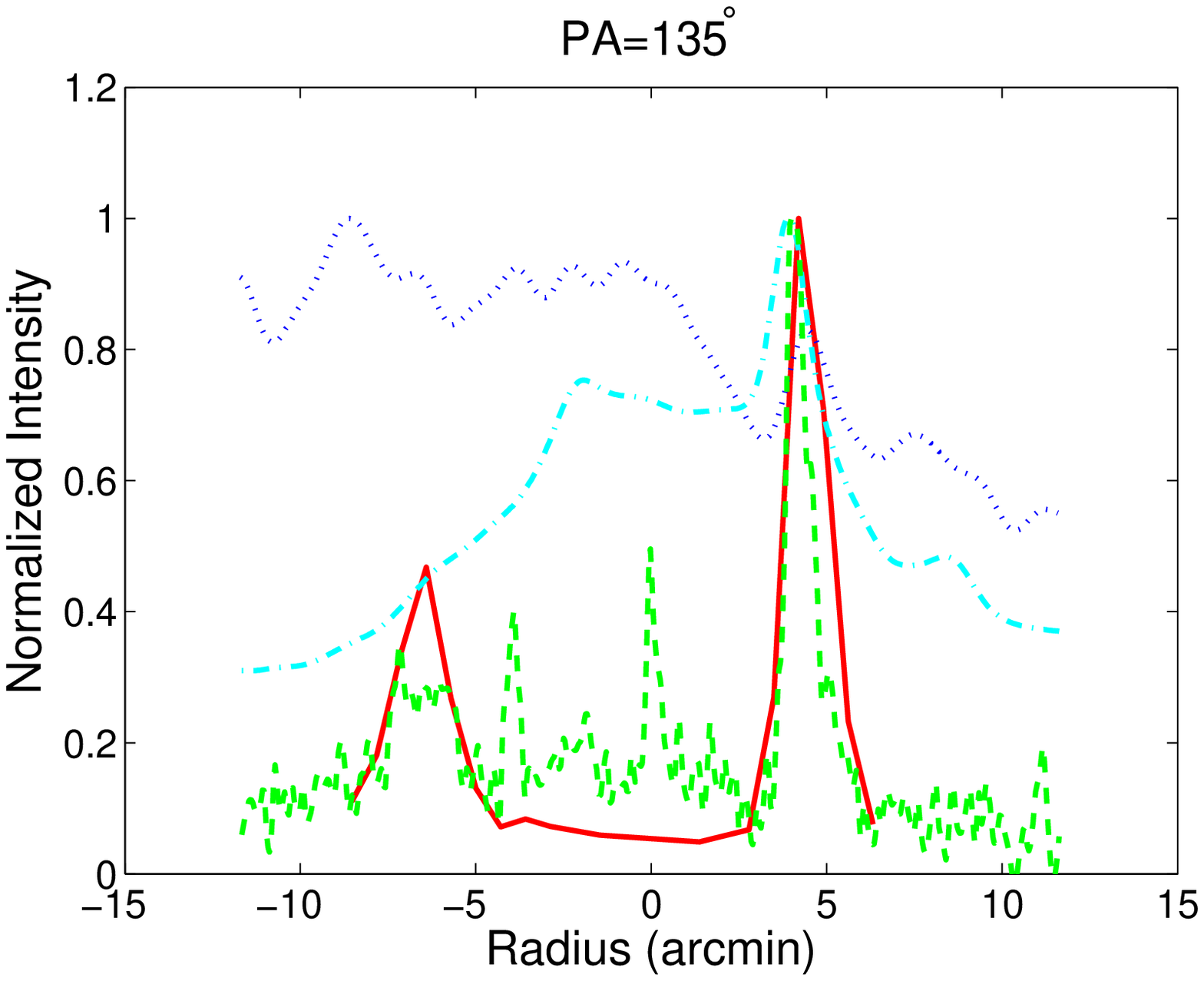}
\end{minipage}
\caption{Normalized intensity distributions of CO, HI, 1420 MHz
radio emission detected in CGPS and 8.3 $\micron$ emission detected
in the MSX A band  along four orientations centered
on the Wolf-Rayet star.}
\end{figure}

\begin{figure}
\includegraphics[angle=0,scale=.80]{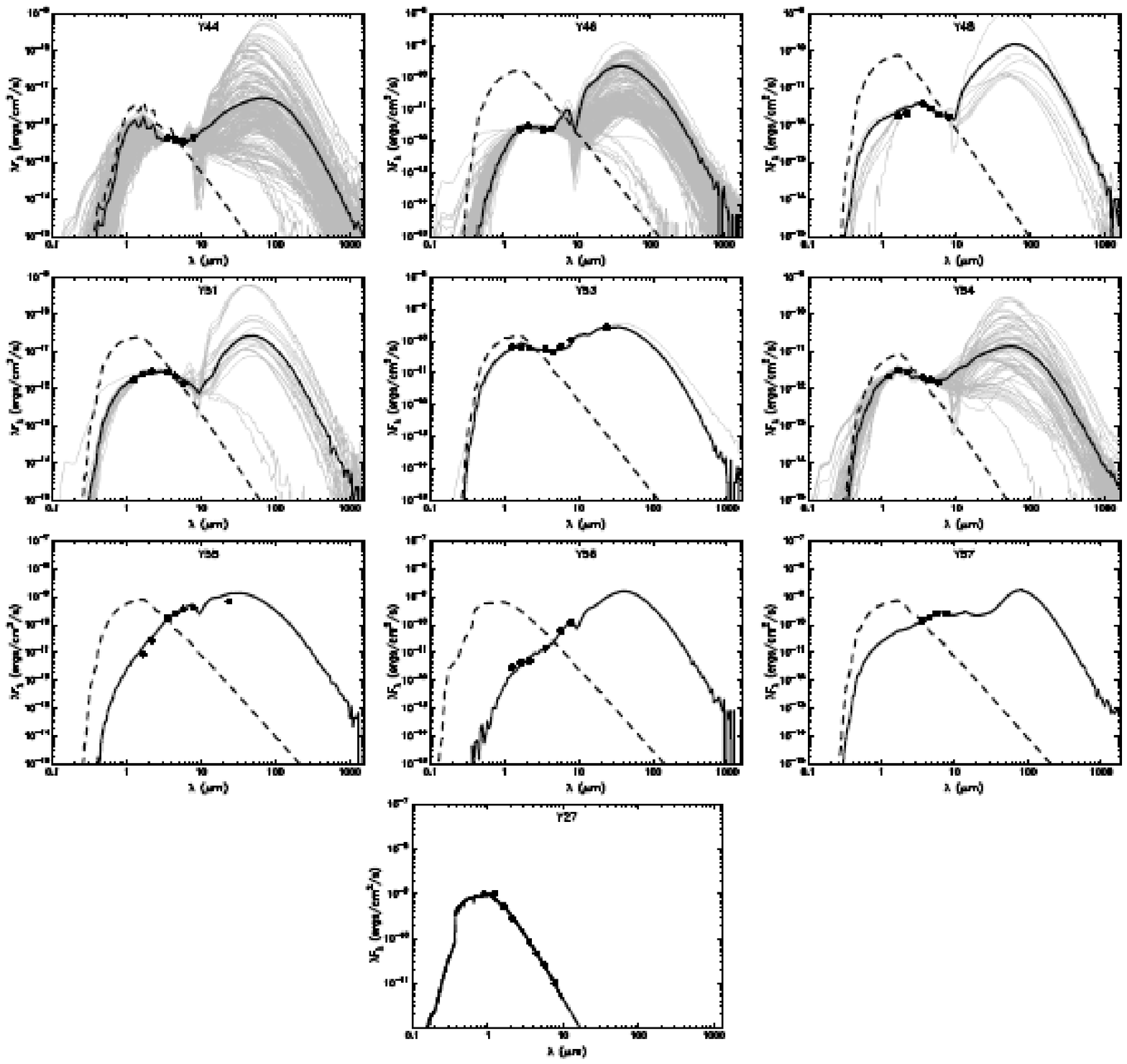}
\centering
\caption{First nine panels present SEDs of the YSOs associated with core A. The filled circles
represent the observed points. The black solid line represents the
best fit to the data, while the gray solid lines represent the fits
with $\chi^{2}-\chi_{best}^{2}<3\times n_{data}$, where $n_{data}$
is the number of data points. The dashed lines represent
photospheric contributions, including the effect of foreground
extinction. The last panel shows the SED of Y27, which favors a stellar photosphere fit rather than a YSO SED fit.}
\end{figure}

\begin{figure}
\begin{minipage}[c]{0.5\textwidth}
  \centering
  \includegraphics[width=65mm,height=85mm,angle=-90]{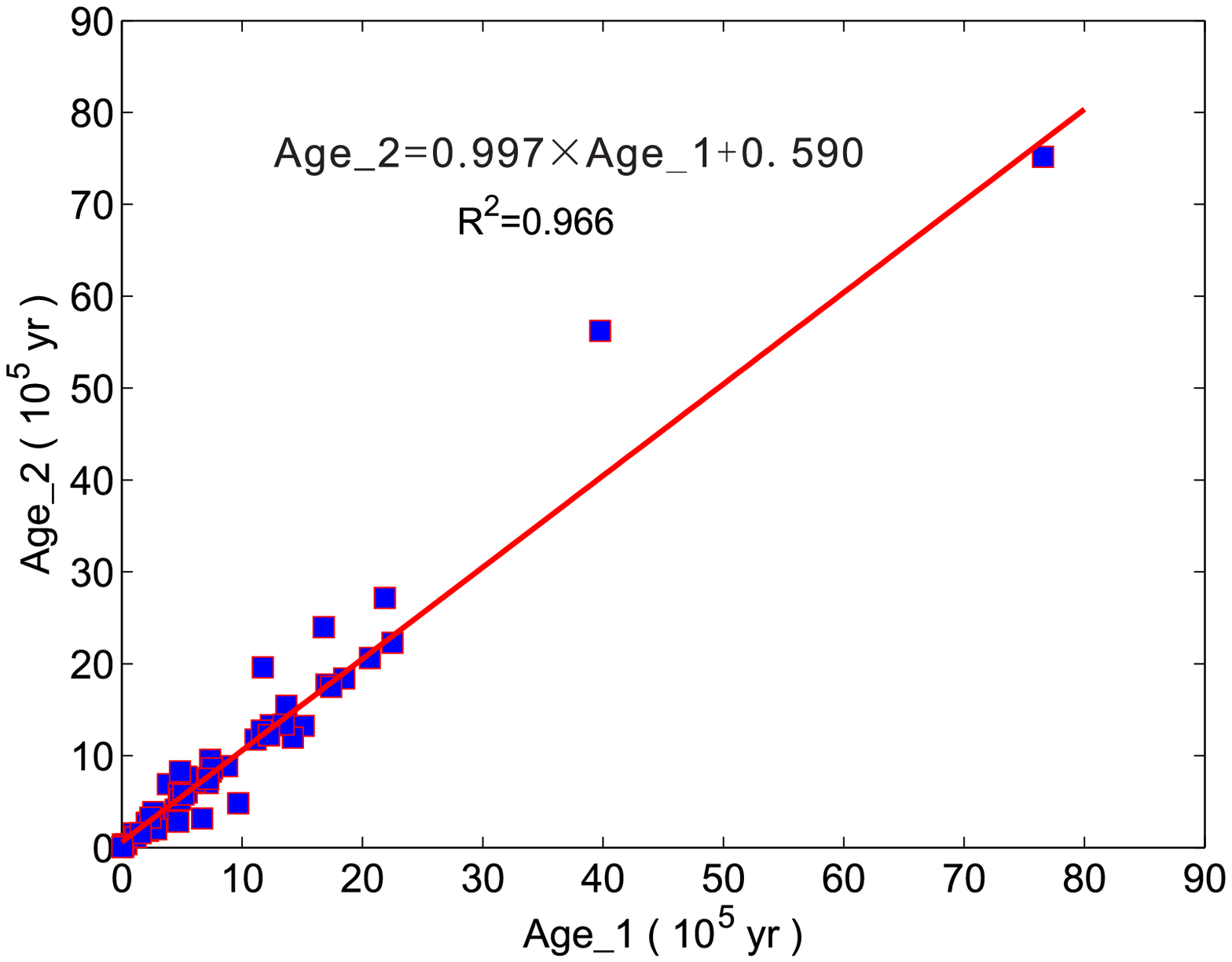}
\end{minipage}
\begin{minipage}[c]{0.5\textwidth}
  \centering
  \includegraphics[width=65mm,height=85mm,angle=-90]{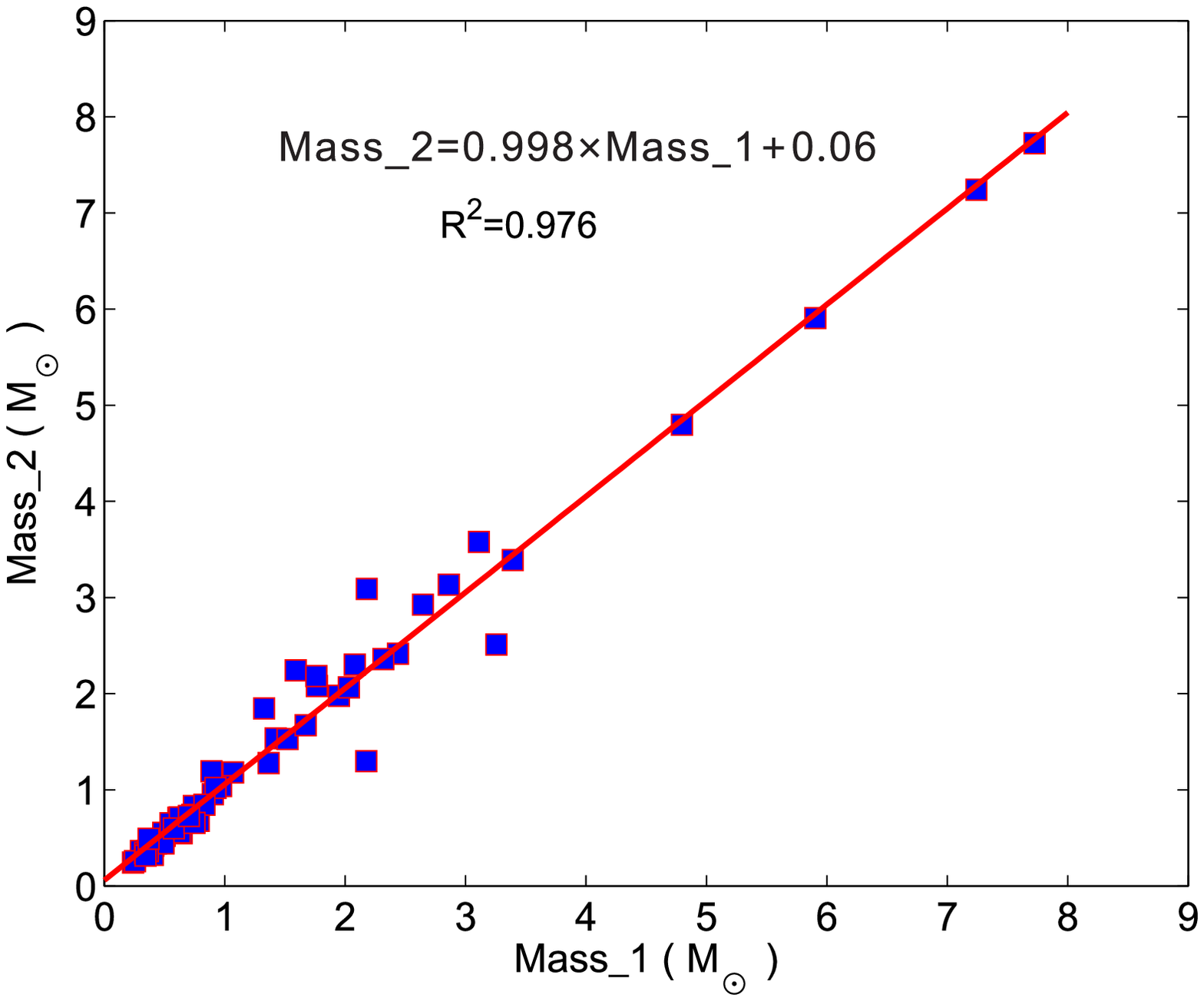}
\end{minipage}
\caption{Comparison of the stellar mass and age of the YSOs in two runnings of the SED modeling. Left: Age comparison; Right: Mass comparison. The solid lines represent the linear fits. Age\_1 and Mass\_1 are the stellar age and mass obtained in the first running by exploring A$_{V}$ in [2,2.5] mag.  Age\_2 and Mass\_2 are the stellar age and mass obtained in the second running by exploring A$_{v}$ in [0,2.5] mag. }
\end{figure}

\begin{figure}
\includegraphics[angle=-90,scale=.50]{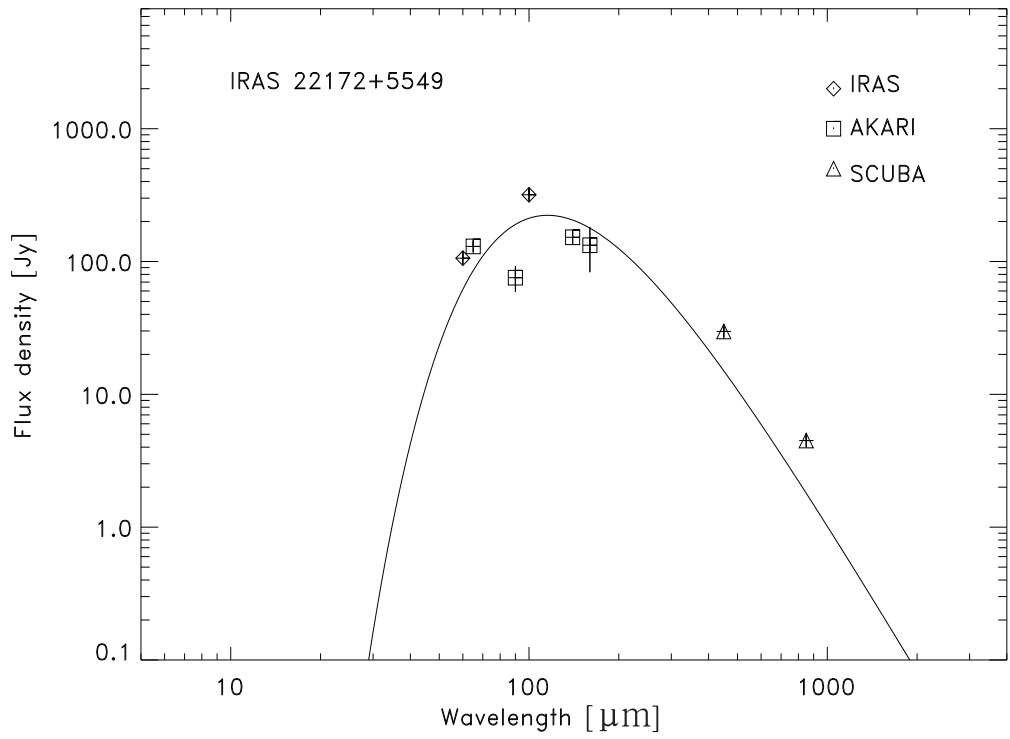}
\caption{SED of molecular core "A" (IRAS 22172+5549). The solid curve represents the best fit. }
\end{figure}

\begin{figure}
\includegraphics[angle=90,scale=.50]{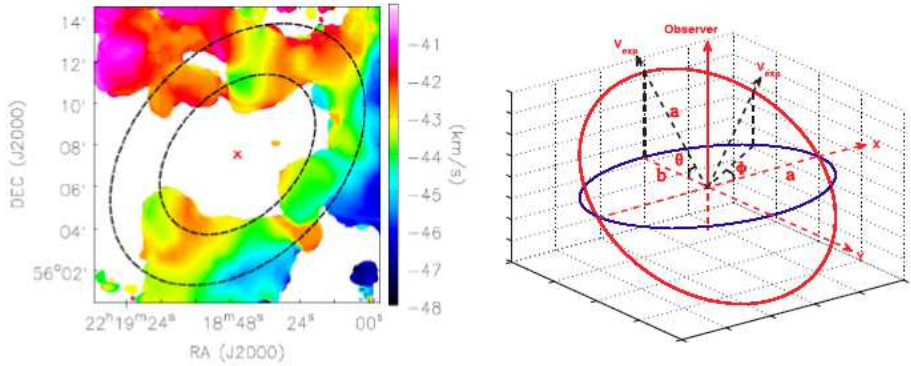}
\caption{Left: first moment map of $^{12}$CO (1-0). Two ellipses are drawn for reference,
to mark the boundaries of the interior and exterior of the molecular ring. Right: Coordinate system for describing the orientation in the sky of the ring nebular.
The axes x and y lie on the plane of the sky. The WR star HD 211853 is located at the origin. The ring nebular is shown as a red circle. The projection of the ring nebular in the sky plane is shown as a blue ellipse. The lengths of the semimajor and semiminor axis of the ellipse are denoted as "a" and "b". The angle $\theta$ is the inclination angle between the plane of the ring nebular and the sky plane. The angle $\phi$ is the position angle measured from the major axis.  }
\end{figure}

\begin{figure}
\includegraphics[angle=90,scale=.50]{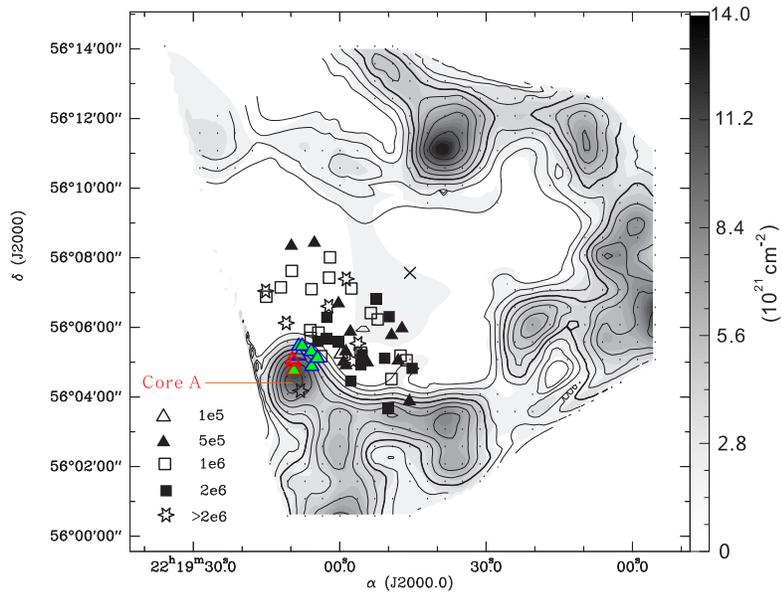}
\caption{T$_{ex}$ in contours overlaid on the column density of
H$_{2}$ in gray scale. The contour levels are from 20\% to 90\% of
the peak (24 K). The young stars in different age intervals are denoted as
different markers on the plot as shown in the lower-left corner. The
upper age of each interval is shown in the lower-left corner. The nine YSOs associated with core "A" are shown as color markers. The three most massive YSOs are shown as markers with red edges. The other six less massive ones are presented as markers with blue edges.}
\end{figure}

\begin{figure}
\includegraphics[angle=0,scale=.50]{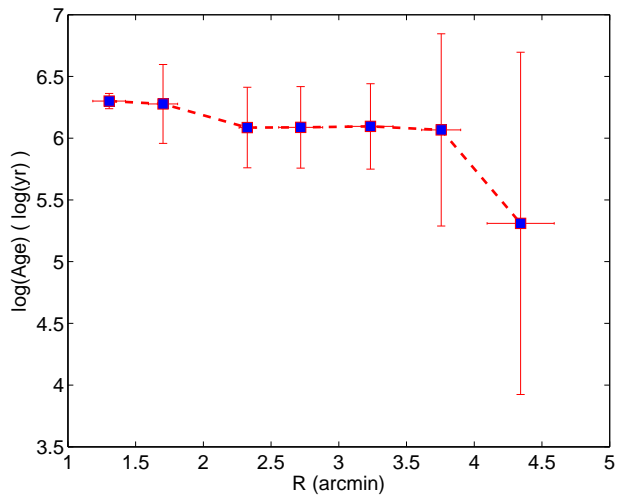}
\caption{Distribution of $0\arcmin.5$ bin averaged ages of YSOs located between HD 211853 and core "A" as a function of radii from the WR star. The errorbars represent the standard deviation in each bin.}
\end{figure}

\clearpage

\begin{deluxetable}{ccrrrrrrrrrrrrrrrcrl}
\tabletypesize{\scriptsize} \tablecolumns{13} \tablewidth{0pc}
\tablecaption{Observed parameters of $^{13}$CO cores} \tablehead{
 \colhead{Core} & \colhead{RA (J2000)}
&\colhead{DEC (J2000)} & \colhead{R\tablenotemark{a}} &\colhead{V$_{lsr}$} &\colhead{T$_{A}$($^{13}$CO)} &\colhead{$\Delta$V($^{13}$CO)} &\colhead{T$_{A}$($^{12}$CO)} &\colhead{$\Delta$v($^{12}$CO)} \\
\colhead{}  & \colhead{(h~m~s)} &
\colhead{($\arcdeg~\arcmin~\arcsec$)} &\colhead{(pc)} & \colhead{(km~s$^{-1}$)} &
\colhead{(K)} & \colhead{(km~s$^{-1}$)}  & \colhead{(K)} &
\colhead{(km~s$^{-1}$)} } \startdata
A  & 22:19:09.4&56:04:46.5 & 1.0 & -43.8(0.3) &2.6(0.1)  &2.4(0.2) & 9.2(0.3) &3.4(0.2) \\
B  & 22:18:38.4&56:03:25.4 & 1.4 & -44.2(0.1) &2.2(0.1)  &2.2(0.1) & 7.5(0.1) &2.7(0.2) \\
C  & 22:18:20.1&56:06:00   & 0.9 & -43.7(0.2) &1.2(0.1)  &2.5(0.1) & 8.5(0.2) &2.6(0.1)  \\
D  & 22:18:03.1&56:08:07.6 & 1.0 & -43.3(0.2) &2.5(0.1)  &1.7(0.2) & 7.6(0.3) &2.6(0.2)  \\
E  & 22:18:03.1&56:05:42.7 & 0.8 & -46.1(0.2) &1.8(0.2)  &2.0(0.2) & 7.9(0.2) &2.5(0.2) \\
F  & 22:18:08.8&56:11:17.3 & 1.1 & -42.7(0.1) &2.1(0.2)  &1.9(0.2) & 7.7(0.3) &2.8(0.2) \\
G  & 22:18:39.1&56:11:17.3 & 1.3 & -43.3(0.1) &2.9(0.1)  &2.6(0.2) &10.1(0.2) &3.5(0.1) \\
H  & 22:19:26.6&56:11:17.3 & 0.5 & -42.0(0.1) &1.0(0.1)  &1.3(0.2) &3.7(0.2)  &2.1(0.1) \\
I  & 22:18:17.6&56:12:23.4 & 1.0 & -50.1(0.1) &1.4(0.1)  &1.1(0.1) &4.9(0.1)  &1.6(0.1) \\
\enddata
\tablenotetext{a}{Taking the distance as 3 kpc}
\end{deluxetable}

\clearpage

\begin{deluxetable}{cccccccccccccccccccccccccccccccccccccccc}
\rotate \tabletypesize{\scriptsize} \tablecolumns{18}
\tablewidth{0pc} \setlength{\tabcolsep}{0.025in}
\tablecaption{Derived parameters of $^{13}$CO cores} \tablehead{
 &\multicolumn{5}{c}{LTE} &\multicolumn{8}{c}{non-LTE}\\
\cline{2-6} \cline{8-15}
\colhead{Core}  &\colhead{$\tau$ ($^{13}$CO)}  &\colhead{T$_{ex}$}&\colhead{N$_{H_{2}}$} &n$_{H_{2}}$
&\colhead{M$_{core}$} & &\colhead{T$_{kin}$} & N$_{H_{2}}$  &n$_{H_{2}}$ &T$_{ex}$ ($^{12}$CO)
& $\tau$ ($^{12}$CO)& T$_{ex}$ ($^{13}$CO) & $\tau$ ($^{13}$CO) &\colhead{M$_{core}$}
&\colhead{M$_{vir}$} &\colhead{M$_{J}$} &\colhead{M$_{pv}$}  &\colhead{P$_{ext}$ / k} &\colhead{P$_{mol}$ / k}\\
\colhead{}  & \colhead{} & \colhead{(K)} &
\colhead{(10$^{21}$~cm$^{-2}$)}& \colhead{(10$^{3}$~cm$^{-3}$)}
&\colhead{(M$_{\sun}$)} &  & \colhead{(K)}  &
\colhead{(10$^{21}$~cm$^{-2}$)} & \colhead{(10$^{3}$~cm$^{-3}$)}  &
\colhead{(K)} & \colhead{} & \colhead{(K)} & \colhead{}
&\colhead{(M$_{\sun}$)} &\colhead{(M$_{\sun}$)} &\colhead{(M$_{\sun}$)} &\colhead{(M$_{\sun}$)}& \colhead{(10$^{6}$ cm$^{-3}$K)} & \colhead{(10$^{5}$ cm$^{-3}$K)}
}\startdata
A    & 0.3 &22   & 10.9  &1.8  &501   & &24.0(0.8)  & 9.0(0.4) &1.4(0.1)  &21.7(0.5) & 10.6(0.8)   &15.6(0.5) &0.5   &410(22)  &504 & 544 &199 & 5.8  &	5.1      \\
B    & 0.3 &18.5 & 7.3   &0.8  &661   & &20.4(0.3)  & 7.8(0.3) &0.9       &18.5(0.2) & 15.6(0.7)   &10.8(0.2) &0.9   &701(30)  &647 & 523 &174 & 3.8  &	2.8    \\
C    & 0.2 &20.5 & 5.2   &0.9  &193   & &26.5(1.1)  & 3.9(0.2) &0.7       &20.5(0.4) &  7.1(0.5)   & 9.9(0.4) &0.5   &145(10)  &473 & 870 &703 & 0.6  &	2.8    \\
D    & 0.3 &17.8 & 7.0   &1.1  &322   & &20.4(0.9)  & 6.9(0.3) &1.1(0.1)  &18.6(0.5) & 14.3(1.2)   &12.2(0.3) &0.8   &319(16)  &357 & 232 &150 & 0.6  &	2.1    \\
E    & 0.2 &19   & 4.3   &0.9  &127   & &22.1(0.8)  & 5.0(0.5) &1.0(0.1)  &19.3(0.4) & 10.3(1.1)   &11.6(0.6) &0.6   &149(16)  &336 & 383 &288 & 0.6  &	2.6   \\
F    & 0.3 &18.8 & 6.4   &0.9  &356   & &21.7(1.0)  & 6.1(0.6) &0.9(0.1)  &18.9(0.5) & 11.5(1.2)   &11.1(0.5) &0.8   &339(323) &439 & 350 &235 & 0.6  &	2.1     \\
G    & 0.3 &23.6 & 13.7  &1.7  &1061  & &26.2(0.6)  &11.3(0.4) &1.4(0.1)  &23.6(0.4) & 11.2(0.6)   &16.7(0.4) &0.6   &877(36)  &710 & 687 &631 & 1.1  &	6.0    \\
H    & 0.3 &11.2 & 1.8   &0.6  &21    & &11.7(0.6)  & 2.4(0.2) &0.8(0.1)  &10.8(0.4) & 15.8(1.9)   & 6.5(0.2) &0.9   & 27(3)   &137 & 121 &51  & 0.6  &	0.9  \\
I    & 0.3 &13.2 & 2.2   &0.3  &99    & &14.6(0.3)  & 3.2(0.2) &0.5       &13.2(0.2) & 19.9(1.2)   & 7.0(0.2) &1.3   &149(10)  &231 & 103 &27  & 0.6  &	0.4   \\
\enddata
\end{deluxetable}

\clearpage

\begin{deluxetable}{ccrrrrrrrrrrrrrrrcrl}
\tabletypesize{\scriptsize} \tablecolumns{13} \tablewidth{0pc}
\tablecaption{Parameters of NVSS radio sources} \tablehead{
 \colhead{Number}& \colhead{name} & \colhead{RA (J2000)}
&\colhead{DEC (J2000)} &\colhead{S$_{1.4}$} &\colhead{a} &\colhead{b} &\colhead{$\theta_{R}$} &\colhead{n$_{e}$} &\colhead{EM}&\colhead{M$_{HII}$}\\
\colhead{}  & \colhead{} & \colhead{(h~m~s)} &
\colhead{($\arcdeg~\arcmin~\arcsec$)} & \colhead{(mJy)} &
\colhead{($\arcsec$)} & \colhead{($\arcsec$)} &
\colhead{($\arcmin$)} &\colhead{(cm$^{-3}$)} &\colhead{(pc
cm$^{-6}$)}&\colhead{(M$_{\sun}$)} } \startdata
1 & 221840+560824 & 22:18:40.77 & +56:08:24.1 &   32.9$\pm$4.1 &   125.9  &   110.0 &1.0&34  &  1322    &1.6     \\
2 & 221846+560534 & 22:18:46.86 & +56:05:34.6 &    4.0$\pm$0.5 &$<$80.5  & $<$43.7  &$<$0.5&$<$33  &    $<$633  &$<$0.2     \\
3 & 221856+560954 & 22:18:56.55 & +56:09:54.6 &   20.3$\pm$2.5 &   83.7  &    65.7  &0.6&54  &  2055    &0.6     \\
4 & 221852+560404 & 22:18:52.33 & +56:04:04.0 &  103.4$\pm$4.1 &   103.0  &   56.1  &0.6&117 &  9961    &1.5     \\
5 & 221905+560442 & 22:19:05.13 & +56:04:42.5 &   87.6$\pm$3.4 &   98.6  &    29.3  &0.4&181 &  16879 & 0.8   \\
6 & 221826+561307 & 22:18:26.97 & +56:13:07.7 &   10.0$\pm$1.4 &   71.7  & $<$50.6  &$<$0.5&$<$52  &    $<$1534     &$<$0.3     \\
7 & 221900+560045 & 22:19:00.72 & +56:00:45.0 &   18.0$\pm$1.8 &   52.4  &    47.5  &0.4&92  &  4026    &0.3     \\
8 & 221929+561202 & 22:19:29.67 & +56:12:02.9 &   15.2$\pm$1.0 &   27.5  & $<$31.8  &$<$0.2&$<$185 &    $<$9675     &$<$0.1     \\
9 & 221912+560045 & 22:19:12.79 & +56:00:45.0 &   38.2$\pm$2.7 &   83.2  &    64.5  &0.6&75  &  3962    &0.9     \\
\enddata
\end{deluxetable}

\clearpage

\begin{deluxetable}{ccccccccccccccccccccccccc}
\tabletypesize{\scriptsize} \setlength{\tabcolsep}{0.05in} \rotate
\tablecaption{The SED fitting results of the YSOs.}
 \tablewidth{0pt} \tablehead{
  Name & \colhead{RA (J2000)}
&\colhead{DEC (J2000)} & Number & $\chi^{2}$ & log(d) &A$_{\nu}$&log(Age) &M$_{*}$ &L$_{*}$  &log(M$_{env}$) & log($\dot{M}_{env}$) & log(M$_{disk})$ & log($\dot{M}_{disk}$) & Stage\\
& \colhead{(h~m~s)} & \colhead{($\arcdeg~\arcmin~\arcsec$)} & & & log(kpc)&(mag)&log(yr)
& M$_{\sun}$  & L$_{\sun}$ & log(M$_{\sun}$)
&log(M$_{\sun}$yr$^{-1}$) & log(M$_{\sun}$)
&log(M$_{\sun}$yr$^{-1}$) &} \startdata
Y1   & 22:18:45.18 &  +56:04:49.3  &35&20.8(1.5)&0.48&2.33(0.04)&6.07(0.08)&1.54(0.20)&8(2)&-1.36(0.42)&-5.72(0.35)&-1.90(0.11)&-7.13(0.11)&0/I\\
Y2   & 22:18:45.73 &  +56:03:52.6  &100&12.3(1.0)&0.48(0.01)&2.29(0.06)&5.43(0.45)&1.18(0.26)&20(9)&-0.62(0.34)&-5.02(0.15)&-1.76(0.22)&-6.20(0.41)&0/I\\
Y3   & 22:18:46.31 &  +56:05:03.4  &306&10.5(0.8)&0.47(0.01)&2.26(0.07)&5.88(0.24)&0.44(0.14)&1&-2.00(0.32)&-5.93(0.31)&-2.83(0.29)&-8.56(0.39)&0/I\\
Y4   & 22:18:47.26 &  +56:05:58.4  &37&12.6(1.7)&0.47(0.01)&2.28(0.06)&5.40(0.10)&2.08(0.36)&37(18)&-0.56(0.26)&-4.71(0.34)&-1.95(0.27)&-6.68(0.31)&0/I\\
Y5   & 22:18:47.54 &  +56:05:11.2  &107&8.0(1.1)&0.47(0.01)&2.27(0.08)&5.95(0.33)&0.24(0.07)&1&-1.26(0.62)&-5.48(0.37)&-2.52(0.20)&-7.64(0.42)&0/I\\
Y6   & 22:18:48.01 &  +56:05:02.9  &1&58.1&0.46&2.00&5.18&1.98&31&-1.08&-4.74&-1.35&-6.75&0/I\\
Y7   & 22:18:49.37 &  +56:05:46.5  &1&130.1&0.46&2.46&5.54&1.30&17&-1.11&-5.75&-1.20&-5.88&0/I\\
Y8   & 22:18:49.40 &  +56:04:31.0  &1172&5.6(1.4)&0.47(0.01)&2.28(0.10)&5.86(0.42)&0.47(0.28)&4(7)&-0.61(1.36)&-4.94(1.00)&-2.36(0.54)&-7.25(0.97)&0/I\\
Y9   & 22:18:49.86 &  +56:06:18.5  &879&14.3(0.6)&0.48(0.01)&2.25(0.06)&6.26(0.12)&1.04(0.16)&2(1)&-2.05(0.65)&-6.14(0.58)&-2.59(0.31)&-8.22(0.82)&III\\
Y10  & 22:18:50.13 &  +56:03:40.8  &24&11.7(1.8)&0.47(0.01)&2.20(0.06)&6.29(0.18)&2.41(0.35)&28(8)&-0.73(0.28)&-4.99(0.24)&-1.49(0.17)&-6.69(0.31)&0/I\\
Y11  & 22:18:50.78 &  +56:05:06.8  &835&5.1(1.4)&0.48(0.01)&2.26(0.10)&6.12(0.29)&0.52(0.19)&1(1)&-1.77(0.60)&-5.79(0.54)&-2.66(0.33)&-8.01(0.64)&0/I\\
Y12  & 22:18:52.22 &  +56:06:13.7  &306&7.4(1.1)&0.48(0.01)&2.25(0.08)&5.78(0.40)&0.55(0.25)&4(4)&-1.09(0.50)&-5.24(0.35)&-2.42(0.24)&-7.64(0.25)&0/I\\
Y13  & 22:18:52.49 &  +56:06:48.8  &550&12.3(1.1)&0.48(0.01)&2.28(0.06)&6.06(0.16)&0.71(0.15)&2(3)&-1.34(1.35)&-5.43(1.83)&-2.37(0.28)&-7.86(1.87)&0/I\\
Y14\tablenotemark{a}  & 22:18:52.69 &  +56:06:05.1  &1&30154.2&0.51&2.00&6.09&3.39&36&-2.98&-8.68&-2.20&-7.59&III\\
Y15  & 22:18:53.62 &  +56:06:24.8  &465&15.0(1.2)&0.48(0.01)&2.34(0.05)&5.92(0.14)&0.64(0.12)&2&-1.46(0.58)&-5.63(0.42)&-2.33(0.20)&-8.14(0.24)&0/I\\
Y16  & 22:18:54.35 &  +56:04:59.6  &81&6.9(1.3)&0.48(0.01)&2.29(0.08)&5.26(0.70)&0.42(0.14)&4(2)&-0.70(0.24)&-4.96(0.20)&-2.14(0.29)&-6.00(0.47)&0/I\\
Y17  & 22:18:55.39 &  +56:05:12.9  &259&5.1(1.5)&0.48(0.01)&2.26(0.10)&6.12(0.30)&0.95(0.30)&3(2)&-1.06(0.78)&-5.30(0.57)&-2.40(0.30)&-7.81(0.36)&0/I\\
Y18  & 22:18:55.67 &  +56:05:16.3  &44&7.6(1.3)&0.47(0.01)&2.22(0.07)&6.08(0.24)&1.53(0.42)&11(7)&-0.55(0.44)&-4.91(0.39)&-2.07(0.19)&-7.37(0.30)&0/I\\
Y19  & 22:18:55.67 &  +56:04:55.4  &2584&1.7(1.9)&0.47(0.02)&2.22(0.16)&5.98(0.57)&0.36(0.28)&1(1)&-1.89(2.11)&-5.87(0.83)&-2.91(0.89)&-8.45(4.77)&0/I\\
Y20\tablenotemark{a}  & 22:18:56.14 &  +56:05:48.6  &1&2131.5&0.48&2.00(0.01)&5.61&4.42&63&-1.10&-5.32&-2.86&-9.64&0/I\\
Y21  & 22:18:56.27 &  +56:05:32.3  &894&7.1(1.2)&0.48(0.01)&2.27(0.08)&6.19(0.23)&0.83(0.25)&2(1)&-1.43(0.59)&-5.56(0.48)&-2.46(0.31)&-8.06(0.39)&0/I\\
Y22  & 22:18:57.18 &  +56:05:02.0  &1&162.5&0.48&2.00&6.38&3.13&128&-2.63&&-7.27&-13.27&III\\
Y23  & 22:18:57.62 &  +56:07:07.0  &43&46.7(0.8)&0.48&2.32(0.03)&5.84(0.08)&1.19(0.10)&6(1)&-1.22(0.14)&-5.34(0.20)&-1.99(0.10)&-7.54(0.09)&0/I\\
Y24  & 22:18:57.71 &  +56:04:27.5  &1084&4.5(1.7)&0.48(0.01)&2.28(0.10)&6.11(0.35)&0.45(0.24)&2(5)&-1.05(2.14)&-5.28(1.84)&-2.50(0.36)&-7.52(1.26)&0/I\\
Y25  & 22:18:57.82 &  +56:05:52.6  &270&16.0(0.8)&0.47(0.01)&2.27(0.06)&5.34(0.47)&0.44(0.18)&5(5)&-0.42(0.65)&-4.90(0.50)&-2.29(0.39)&-6.84(1.40)&0/I\\
Y26  & 22:18:58.38 &  +56:04:23.3  &24&15.1(1.4)&0.48&2.34(0.06)&6.88(0.04)&2.24(0.28)&46(29)&0.16(0.39)&-6.38(0.39)&-2.21(0.25)&-7.46(0.37)&III\\
Y27\tablenotemark{b}  & 22:18:58.68 &  +56:07:23.6  &1&19476.0&0.51&2.50&6.11&14.60&19200&-7.86&&-8.04&-9.44&III\\
Y28  & 22:18:58.71 &  +56:04:54.5  &53&5.3(1.6)&0.48(0.01)&2.24(0.09)&5.50(0.76)&0.55(0.39)&6(8)&-0.55(0.95)&-4.99(0.75)&-2.41(0.67)&-7.32(0.35)&0/I\\
Y29  & 22:18:58.76 &  +56:05:18.8  &1106&4.6(1.6)&0.47(0.01)&2.26(0.10)&5.55(0.38)&0.26(0.15)&1(1)&-1.61(0.67)&-5.60(0.62)&-2.85(0.42)&-8.17(1.60)&0/I\\
Y30  & 22:18:59.21 &  +56:05:00.6  &62&2.9(2.1)&0.47(0.01)&2.27(0.10)&5.46(0.77)&0.56(0.25)&3(4)&-1.07(0.43)&-5.23(0.51)&-2.19(0.34)&-7.95(0.56)&0/I\\
Y31  & 22:18:59.72 &  +56:05:13.9  &1&128.7&0.49&2.02&4.58&1.67&33&-0.33&-4.82&-1.24&-6.01&0/I\\
Y32  & 22:19:00.08 &  +56:03:14.8  &314&18.4(0.7)&0.48(0.01)&2.31(0.04)&6.88(0.02)&2.36(0.06)&31(3)&-3.63(0.21)&&-3.06(0.29)&-7.99(0.52)&III\\
Y33  & 22:19:00.18 &  +56:05:34.7  &55&6.0(1.5)&0.47(0.01)&2.22(0.08)&6.01(0.36)&0.32(0.11)&2(1)&-1.17(0.39)&-5.29(0.40)&-2.35(0.21)&-7.26(0.43)&0/I\\
Y34  & 22:19:00.26 &  +56:06:41.4  &178&4.4(1.6)&0.48(0.01)&2.22(0.09)&5.59(0.57)&0.71(0.35)&6(8)&-0.72(0.60)&-5.03(0.42)&-2.36(0.35)&-7.50(0.56)&0/I\\
Y35  & 22:19:02.01 &  +56:07:08.5  &7&22.0(1.2)&0.47&2.15(0.04)&6.43(0.09)&2.30(0.11)&12(2)&-1.65(0.23)&-5.96(0.24)&-1.85(0.19)&-7.75(0.24)&0/I\\
Y36  & 22:19:02.01 &  +56:08:00.7  &1391&7.8(1.1)&0.48(0.01)&2.27(0.08)&5.85(0.33)&0.34(0.16)&1(2)&-1.32(1.17)&-5.48(1.20)&-2.64(0.34)&-7.68(1.05)&0/I\\
Y37  & 22:19:02.18 &  +56:07:25.8  &1174&3.7(1.7)&0.48(0.01)&2.28(0.11)&5.98(0.41)&0.65(0.33)&2(2)&-1.25(0.74)&-5.41(0.61)&-2.44(0.37)&-7.78(0.39)&0/I\\
Y38  & 22:19:02.26 &  +56:06:36.5  &320&3.7(1.7)&0.48(0.01)&2.28(0.10)&6.25(0.26)&1.02(0.34)&3(2)&-1.25(1.46)&-5.58(0.97)&-2.41(0.46)&-8.06(0.43)&0/I\\
Y39  & 22:19:02.62 &  +56:06:17.3  &706&4.2(1.7)&0.47(0.01)&2.26(0.11)&6.13(0.33)&0.44(0.25)&2(5)&-1.05(2.41)&-5.23(2.22)&-2.61(0.42)&-7.79(1.71)&0/I\\
Y40  & 22:19:02.77 &  +56:05:40.6  &116&14.9(1.2)&0.48(0.01)&5.03(0.53)&6.27(0.12)&1.85(0.27)&20(10)&-1.25(0.60)&-5.66(0.48)&-2.27(0.18)&-8.16(0.31)&0/I\\
Y41  & 22:19:03.80 &  +56:05:10.3  &81&12.1(1.2)&0.47(0.01)&2.27(0.06)&5.85(0.36)&0.68(0.22)&4(2)&-1.04(0.46)&-5.26(0.28)&-2.07(0.13)&-7.21(0.24)&0/I\\
Y42  & 22:19:04.36 &  +56:05:50.8  &395&7.1(1.2)&0.48(0.01)&2.29(0.08)&5.94(0.41)&0.74(0.36)&11(12)&0.00(0.86)&-4.59(0.56)&-2.04(0.48)&-6.49(0.77)&0/I\\
Y43  & 22:19:04.46 &  +56:05:36.0  &1&113.3&0.48&2.42&6.09&3.39&36&-2.98&-8.68&-2.20&-7.59&III\\
Y44  & 22:19:04.50 &  +56:05:07.8  &215&6.3(1.4)&0.47(0.01)&2.26(0.09)&5.62(0.52)&0.68(0.34)&10(10)&-0.32(0.70)&-4.72(0.58)&-2.19(0.43)&-6.68(1.48)&0/I\\
Y45  & 22:19:05.21 &  +56:08:25.8  &10&46.4(0.9)&0.49&2.31(0.03)&5.42(0.04)&0.85(0.08)&7(1)&-0.66(0.14)&-4.92(0.14)&-2.01(0.07)&-7.22(0.08)&0/I\\
Y46  & 22:19:05.72 &  +56:04:52.7  &218&6.7(1.3)&0.48(0.01)&2.27(0.09)&5.30(0.59)&2.07(0.69)&62(37)&0.18(0.60)&-4.60(0.26)&-1.42(0.37)&-6.17(0.36)&0/I\\
Y47  & 22:19:05.79 &  +56:07:05.9  &138&10.9(0.9)&0.48(0.01)&2.29(0.06)&5.87(0.12)&0.68(0.15)&2&-1.87(0.62)&-6.15(0.31)&-2.58(0.22)&-8.21(0.31)&III\\
Y48  & 22:19:05.82 &  +56:05:18.8  &10&108.1(0.6)&0.47&2.29(0.02)&5.06(0.03)&2.51(0.19)&74(13)&0.72(0.06)&-4.08(0.04)&-2.22(0.04)&-7.37(0.08)&0/I\\
Y49  & 22:19:05.94 &  +56:05:41.7  &262&4.9(1.6)&0.48(0.01)&2.27(0.09)&5.89(0.31)&0.71(0.31)&5(8)&-0.74(0.95)&-4.83(1.21)&-2.45(0.75)&-7.32(2.15)&0/I\\
Y50  & 22:19:06.02 &  +56:05:55.3  &1&41.0&0.51&2.00&5.77&3.58&24&-1.46&-5.25&-5.07&-10.84&0/I\\
Y51  & 22:19:07.65 &  +56:05:28.6  &22&14.5(1.6)&0.48(0.01)&2.23(0.06)&5.68(0.30)&3.09(0.40)&67(25)&-0.42(0.13)&-4.60(0.15)&-1.61(0.27)&-6.47(0.42)&0/I\\
Y52  & 22:19:08.03 &  +56:04:11.1  &1&130.6&0.44&2.00&6.31&4.80&408&-5.15&&-0.82&-7.62&III\\
Y53  & 22:19:08.30 &  +56:05:11.0  &2&387.4(0.6)&0.45&2.16(0.01)&4.51(0.02)&2.93(0.03)&98(1)&0.07(0.01)&-4.41(0.01)&-1.14(0.03)&-5.57(0.03)&0/I\\
Y54  & 22:19:08.45 &  +56:05:28.7  &72&9.4(1.8)&0.48(0.01)&2.27(0.07)&5.51(0.35)&1.28(0.43)&18(10)&0.27(0.49)&-4.42(0.27)&-1.61(0.18)&-6.42(0.17)&0/I\\
Y55  & 22:19:09.38 &  +56:05:00.4  &1&299.6&0.46&2.50&4.12&7.72&607&1.19&-4.40&-0.92&-5.96&0/I\\
Y56  & 22:19:09.39 &  +56:04:45.7  &1&193.3&0.44&2.00&5.20&7.24&1003&1.88&-4.04&-0.96&-7.04&0/I\\
Y57  & 22:19:09.70 &  +56:05:04.8  &1&30.7&0.48&2.00&3.61&5.91&478&1.22&-2.92&-2.32&-6.35&0/I\\
Y58  & 22:19:09.84 &  +56:07:37.4  &1438&4.1(1.5)&0.48(0.01)&2.27(0.11)&5.72(0.54)&0.36(0.24)&2(3)&-1.33(0.92)&-5.42(0.53)&-2.55(0.43)&-7.76(0.88)&0/I\\
Y59  & 22:19:09.94 &  +56:08:20.5  &174&8.4(1.1)&0.48(0.01)&2.27(0.07)&5.45(0.40)&0.32(0.14)&2(2)&-1.19(0.36)&-5.28(0.25)&-2.42(0.28)&-7.33(0.52)&0/I\\
Y60  & 22:19:10.94 &  +56:06:07.6  &1134&5.4(1.4)&0.48(0.01)&2.26(0.09)&6.24(0.23)&0.60(0.20)&1&-1.97(0.99)&-6.00(0.68)&-2.77(0.48)&-8.48(0.47)&III\\
Y61  & 22:19:12.07 &  +56:07:09.1  &54&9.8(1.2)&0.47(0.01)&2.20(0.07)&5.76(0.18)&0.66(0.14)&3(2)&-1.12(0.36)&-5.31(0.25)&-2.28(0.37)&-7.15(0.45)&0/I\\
Y62  & 22:19:13.80 &  +56:07:48.4  &305&15.9(1.2)&0.49(0.01)&2.82(0.42)&6.75(0.04)&2.18(0.12)&26(7)&-3.39(0.40)&-8.72(2.28)&-3.04(0.82)&-10.24(0.57)&III\\
Y63  & 22:19:15.06 &  +56:06:53.0  &194&4.0(1.7)&0.48(0.01)&2.24(0.10)&5.92(0.47)&0.49(0.24)&1(1)&-1.62(0.89)&-5.73(0.75)&-2.49(0.48)&-8.20(0.44)&0/I\\
Y64  & 22:19:15.18 &  +56:07:02.8  &2239&2.2(1.7)&0.47(0.02)&2.26(0.14)&6.35(0.29)&0.73(0.35)&1(2)&-1.85(13.16)&-6.11(9.52)&-2.85(0.85)&-8.51(5.91)&III\\
\enddata
\tablenotetext{a}{Bad fit.}
\tablenotetext{b}{Evolved star?}
\end{deluxetable}


\clearpage
\begin{thebibliography}{}
{\small
\bibitem[Beaumont \& Williams(2010)]{bea10}Beaumont, C. N., \& Williams, J. P., 2010, 709, 791

\bibitem[Bertoldi(1989)]{ber89}Bertoldi, F. 1989, \apj, 346, 735

\bibitem[Bertoldi \& McKee(1990)]{ber90}Bertoldi, F. \& McKee, C. F. 1990, \apj, 354, 529

\bibitem[Bisbas et al.(2011)]{bis11}Bisbas, T. G., W\"{u}nsch, R., Whitworth, A. P., Hubber, D. A., Walch, S., 2011, \apj, 736, 142

\bibitem[Brand et al.(2011)]{bra11}Brand, J., Massi, F., Zavagno, A., Deharveng, L., Lefloch, B., 2011, \aap, 527, 62

\bibitem[Cappa et al.(2008)]{cap08}Cappa, C. E., Vasquez, J., Arnal, E. M., Cichowolski, S., Pineault, S., 2008, RMxAC, 33, 142

\bibitem[Chen et al.(2007)]{chen07}Chen, W. P., Lee, H. T., Sanchawala, K., 2007, IAUS, 237, 278

\bibitem[Condon et al.(1998)]{con98}Condon, J. J., Cotton, W. D., Greisen, E. W., Yin, Q. F., Perley, R. A., et al., 1998, \aj, 115, 1693

\bibitem[Dent et al.(2009)]{dent09}Dent, W. R. F., Hovey, G. J., Dewdney, P. E., Burgess, T. A., Willis, A. G., et al., 2009, \mnras, 395, 1805

\bibitem[Deharveng et al.(2003)]{de03}Deharveng, L., Lefloch, B., Zavagno, A., Caplan, J., Whitworth, A. P., et al. 2003, \aap, 408, L25

\bibitem[Deharveng, Zavagno, \& Caplan(2005)]{de05}Deharveng, L., Zavagno, A., \& Caplan, J., 2005, \aap, 433,565

\bibitem[Deharveng et al.(2008)]{de08}Deharveng, L., Lefloch, B., Kurtz, S., Nadeau, D., Pomar\`{e}s, M., et al., 2008, \aap, 482, 585

\bibitem[Di Francesco et al.(2008)]{di08}Di Francesco, J., Johnstone, D., Kirk, H., MacKenzie, T., Ledwosinska, E., 2008, \apjs, 175, 277

\bibitem[Elmegreen \& Lada(1977)]{el77}Elmegreen, B. G., \& Lada, C. J. 1977, \apj, 214, 725

\bibitem[Garden et al.(1991)]{gar91}Garden, R. P., Hayashi, M., Hasegawa, T., Gatley, I., Kaifu,
N., 1991, \apj, 374, 540

\bibitem[Grave \& Kumar(2009)]{gra09}Grave J. M. C. \& Kumar M. S. N., 2009, \aap, 498, 147

\bibitem[Gritschneder et al.(2009)]{gri09}Gritschneder, M., Naab, T., Burkert, A., Walch, S., Heitsch, F., et al., 2009, \mnras, 393, 21

\bibitem[Guilloteau \& Lucas(2000)]{gui00}Guilloteau, S. \& Lucas, R., 2000, in Astronomical Society of the Pacific Conference Series, Vol. 217, Imaging at Radio through Submillimeter
Wavelengths, ed. J. G. Mangum \& S. J. E. Radford, 299

\bibitem[Harten, Felli \& Tofani.(1978)]{har78}Harten R. H., Felli M., Tofani G., 1978, \aap, 70, 205

\bibitem[Haworth \& Harries.(2012)]{haw12}Haworth, T. J. \& Harries, T. J., 2012, \mnras, 420, 562

\bibitem[Hennebelle \& Chabrier(2008)]{hen08}Hennebelle, P., \& Chabrier, G., 2008, \apj, 684, 395

\bibitem[Hester \& Desch(2005)]{he05}Hester, J. J., \& Desch, S. J. 2005, ASP Conf. Ser. 341: Chondrites and the Protoplanetary Disk, 341, 107

\bibitem[Lefloch \& Lazareff.(1994)]{lef94}Lefloch B, \& Lazareff, B., 1994, \aap, 289, 559

\bibitem[Li \& Goldsmith(2003)]{li03}Li, D., \& Goldsmith, P. F. 2003, \apj, 585, 823

\bibitem[Liu et al.(2010)]{liu10}Liu, T., Wu, Y.-F., Wang, K., 2010, RAA, 10, 67

\bibitem[Liu et al.(2011)]{liu11}Liu, T., Zhang, H. W., Wu, Y. F., Qin, S.-L., Miller, M., 2011, \apj, 734, 22

\bibitem[Marston (1996)]{mar96}Marston, A. P., 1996, \aj, 112, 2828

\bibitem[Mart\'{\i}n-Pintado et al.(1999)]{mart99}Mart\'{\i}n-Pintado, J., Gaume, R. A., Rodr\'{\i}guez-Fern\'{a}ndez, N., de Vicente, P., Wilson, T. L., 1999, \apj, 519, 667

\bibitem[Miao et al.(2009)]{miao09}Miao J., White G. J., Thompson M. A., Nelson R. P., 2009, \apj, 692, 382

\bibitem[Morgan et al.(2004)]{mor04}Morgan, L. K., Thompson, M. A., Urquhart, J. S., White, G. J., Miao, J., 2004, \aap, 426, 535

\bibitem[Morgan et al.(2010)]{mor10}Morgan, L. K., Figura, C. C., Urquhart, J. S., Thompson, M. A., 2010, \mnras, 408, 157

\bibitem[Myers, Linke \& Benson(1983)]{my83}Myers, P. C., Linke, R. A., \& Benson, P. J., 1983, \apj, 264, 517

\bibitem[Nishimaki et al.(2008)]{ni08}Nishimaki, Y., Yamamuro, T., Motohara, K., Miyata, T., Tanaka, M., 2008, PASJ, 60, 191

\bibitem[Ogura(2006)]{og06}Ogura, Katsuo., 2006, BASI, 34, 1110

\bibitem[Ogura(2010)]{og10}Ogura, K., 2010, ASInC, 1, 190

\bibitem[Ossenkopf \& Henning(1994)]{oss94}Ossenkopf, V., \& Henning, T. 1994, \aap, 291, 943

\bibitem[Panagia \& Walmsley(1978)]{pan78}Panagia, N. \& Walmsley, C. M., 1978, \aap, 70, 411

\bibitem[Petriella et al.(2010)]{pe10}Petriella, A., Paron, S., \& Giacani, E. 2010, \aap, 513, A44+

\bibitem[Pomar\`{e}s, M. et al.(2009)]{po09}Pomar\`{e}s, M., Zavagno, A., Deharveng, L., Cunningham, M., Jones, P., et al., 2009, \aap, 494, 987

\bibitem[Price et al.(2001)]{pro01}Price, S. D., Egan, M. P., Carey, S. J., Mizuno, D. R., Kuchar, T. A., 2001, \aj, 121, 2819

\bibitem[Qiu et al.(2008)]{qiu08}Qiu, Keping., Zhang, Qizhou., Megeath, S. Thomas., Gutermuth, Robert A., Beuther, Henrik., et al, 2008, \apj, 685, 1005

\bibitem[Robitaille et al.(2006)]{ro06}Robitaille T. P., Whitney B. A., Indebetouw R., Wood K., \& Denzmore P., 2006, \apjs, 167, 256

\bibitem[Robitaille et al.(2007)]{ro07}Robitaille T. P., Whitney B. A., Indebetouw R., and Wood K., 2007, \apjs, 169, 328

\bibitem[Saurin et al.(2010)]{sau10}Saurin, T. A., Bica, E., \& Bonatto, C., 2010, \mnras, 407, 133

\bibitem[Shan et al.(2009)]{shan09}Shan, W., Li, Z., Zhong, J., Shi, S., 2009, IEEE Trans. On Appl. Supercond, 19, 432.

\bibitem[Sharpless(1959)]{shar59}Sharpless S., 1959, \apjs, 4, 257

\bibitem[Sugitani et al.(1991)]{su91}Sugitani, K., Fukui, Y., Ogura, K., 1991, \apjs, 77, 59

\bibitem[Sugitani \& Ogura(1994)]{su94}Sugitani, K., \& Ogura, K., 1994, \apjs, 92, 163

\bibitem[Sun et al.(2006)]{sun06}Sun, K., Kramer, C., Ossenkopf, V., Bensch, F., Stutzki, J., et al., 2006, \aap, 451, 539

\bibitem[Taylor et al.(2003)]{tay03}Taylor, A. R., Gibson, S. J., Peracaula, M., Martin, P. G., Landecker, T. L., et al., 2003, \aj, 125, 3145

\bibitem[Thompson et al.(2004)]{tho04}Thompson, M. A., White, G. J., Morgan, L. K., Miao, J., Fridlund, C. V. M., et al., 2004, \aap, 414, 1017

\bibitem[Ungerechts et al.(2000)]{ung00}Ungerechts H., Umbanhowar P., \& Thaddeus P., 2000, \apj, 537, 221

\bibitem[Urquhart et al.(2004)]{ur04}Urquhart J. S., Thompson M. A., Morgan L. K., White G. J., 2004, \aap, 428, 723

\bibitem[Urquhart et al.(2006)]{ur06}Urquhart J. S., Thompson M. A., Morgan L. K., White G. J., 2006,
\aap, 450, 625

\bibitem[Urquhart et al.(2007)]{ur07}Urquhart J. S., Thompson M. A., Morgan L. K., Pestalozzi M. R.,
White G. J., Muna D. N., 2007, \aap, 467, 1125

\bibitem[Van der Tak et al.(2007)]{van07}Van der Tak, F.F.S., Black, J.H., Sch\"{o}ier, F.L., Jansen, D.J., van Dishoeck, E.F. 2007, \aap, 468, 627

\bibitem[Vasquez et al.(2010)]{vas10}Vasquez, J., Cappa, C. E., Pineault, S., Duronea, N. U., 2010, \mnras, 405, 1976

\bibitem[Wang et al.(2009)]{wang09}Wang, K., Wu, Y. F., Ran, L., Yu, W. T., Miller, M., 2009, \aap, 507, 369

\bibitem[Whitworth et al.(1994)]{wh94}Whitworth, A. P., Bhattal, A. S., Chapman, S. J., Disney, M. J., \& Turner, J. A. 1994, \mnras, 268, 291

\bibitem[Wu et al.(2003)]{wu03}Wu, Y., Wang, J., \& Wu, J., 2003, Chin. Phys. Letter, 20, 1409

\bibitem[Wu et al.(2001)]{wu01}Wu, Y., Wu, J., Wang, J., 2001, \aap, 380, 665

\bibitem[Zavagno et al.(2006)]{za06}Zavagno, A., Deharveng, L., Comer\'{o}n, F., Brand, J., Massi, F., et al., 2006, \aap, 446, 171

\bibitem[Zavagno et al.(2007)]{za07}Zavagno, A., Pomar\`{e}s, M., Deharveng, L., Hosokawa, T., Russeil, D., et al., 2007, \aap, 472, 835

\bibitem[Zhu et al.(2010)]{zhu10}Zhu, L., Wright, M. C. H., Zhao, J-H., Wu, Y. F., 2010, \apj, 712, 674
}
\end{thebibliography}
\end{document}